# Association of a planetary tidal effect with the time variation of the ~13.5 day component of geomagnetic activity.

I. R. Edmonds

12 Lentara St, Kenmore, Brisbane, 4069 Australia.
ian@solartran.com.au

**Abstract**
We show that there is a previously unreported quad-annual variation in the ~13.5 day component of the aa index of geomagnetic activity. We derive a model based on the planetary tidal effect at the solar surface due to Mercury and Jupiter that, when combined with an equinoctial response of the magnetosphere, predicts the times of occurrence of predominantly quad-annual variation or predominantly semi-annual variation in the ~13.5 day component of the aa index. In support of the model we show that, during years when the quad-annual variation in the ~13.5 day component of aa index is predominant there is a large component at the 88 day periodicity of Mercury in the ~13.5 day component of solar wind speed. As further support for the model we establish that significant peaks in the aa index spectrum are due to an 88 day modulation of 27 day period solar activity. The model also predicts the occurrence of planetary tidal effect maximum in anti-phase with solar cycle maximums around 1970 and we show this is consistent with prior observations of higher solar emissions during the 1975 solar cycle minimum than in the following solar cycle maximum. This effect offers a possible explanation of the reduced solar cycle maximums around 1970 and during the Dalton Minimum.

## 1. Introduction.
The interaction of the solar wind and the interplanetary magnetic field with the magnetosphere results in variations in the terrestrial magnetic field. Daily or hourly observations of the variations are reduced to indices of geomagnetic activity the most common being the aa indices. The aa index has been observed from 1868 to the present and has been studied extensively, Mayaud (1972). The principal periodic components of the aa index occur at 11 years (the solar cycle), 182 days (the semi-annual variation), Cliver et al (1996), and the fundamental and harmonics of the solar rotation period at 27, 13.5, 9 and 6.75 days, De Meyer (2006). The semi-annual time variation of geomagnetic activity has maxima near the equinoxes and minima near the solstices and is generally attributed to one or more of three mechanisms: the axial or Rosenberg-Coleman mechanism, the equinoctial mechanism, and the Russell-McPherron mechanism, Cortie (1912), Rosenberg and Coleman (1969), Russell and McPherron (1973), Cliver et al (2002), Cliver et al (2004). The different mechanisms are associated with (a), the deviation of the Earth's orbit to higher solar latitudes around the equinoxes, (b), the angle between the Sun-Earth line and Earth's magnetic dipole or (c), some combination of (a) and (b), Cliver et al (2004). The difference in timing of the semi-annual maximum in the three different mechanisms is only about one month (7 March – 7 April) making it challenging to distinguish between the mechanisms, with some studies indicating that all the different mechanisms apply to some extent in selected years, Cliver et al (2004).

The present work focuses on a, previously unreported, quad-annual variation in the ~13.5 day component of the aa index. The ~13.5 day component was isolated by band pass filtering as described in the section on method and data sources below. In this paper the ~13.5 day component is related to the variation of aa index in the period range 12.5 to 16 days and, in this paper, is labelled as 13.5aa index. A large fraction of variation in the aa index is shared roughly equally between variation at ~27 day period and variation at ~13.5 day period. The combined variations at 182 days (semi-annual), ~9 days and ~6.75 days periods comprise a similar fraction with all of these systematic variations superimposed on a substantial background noise variation, Cliver et al (1996). The quad-annual variation has four maximums each year with minimums at the two solstices and at the two equinoxes. Examples of the quad-annual variation in the ~13.5 day component of aa index are shown in Figures 1A to 1D.

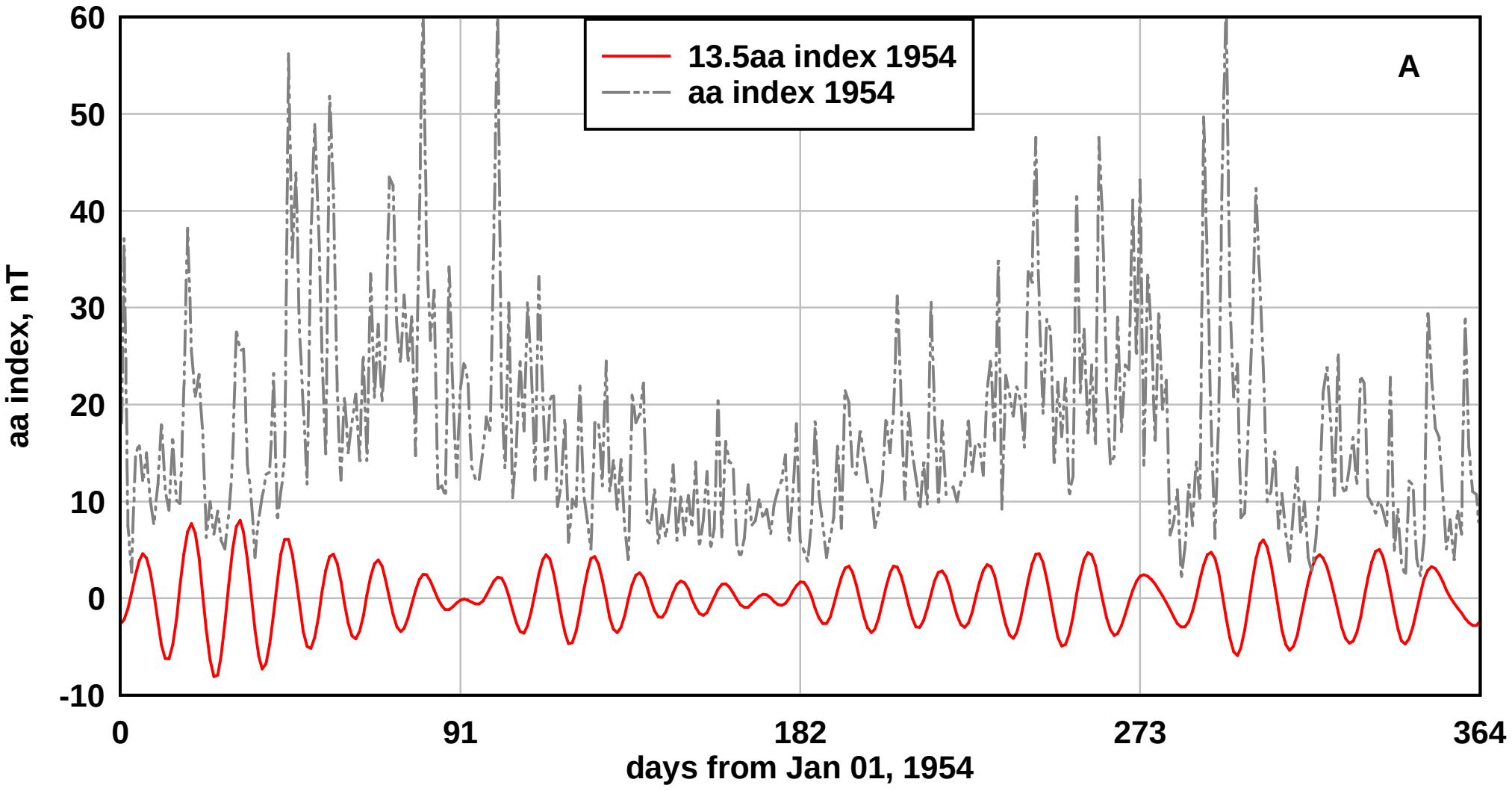

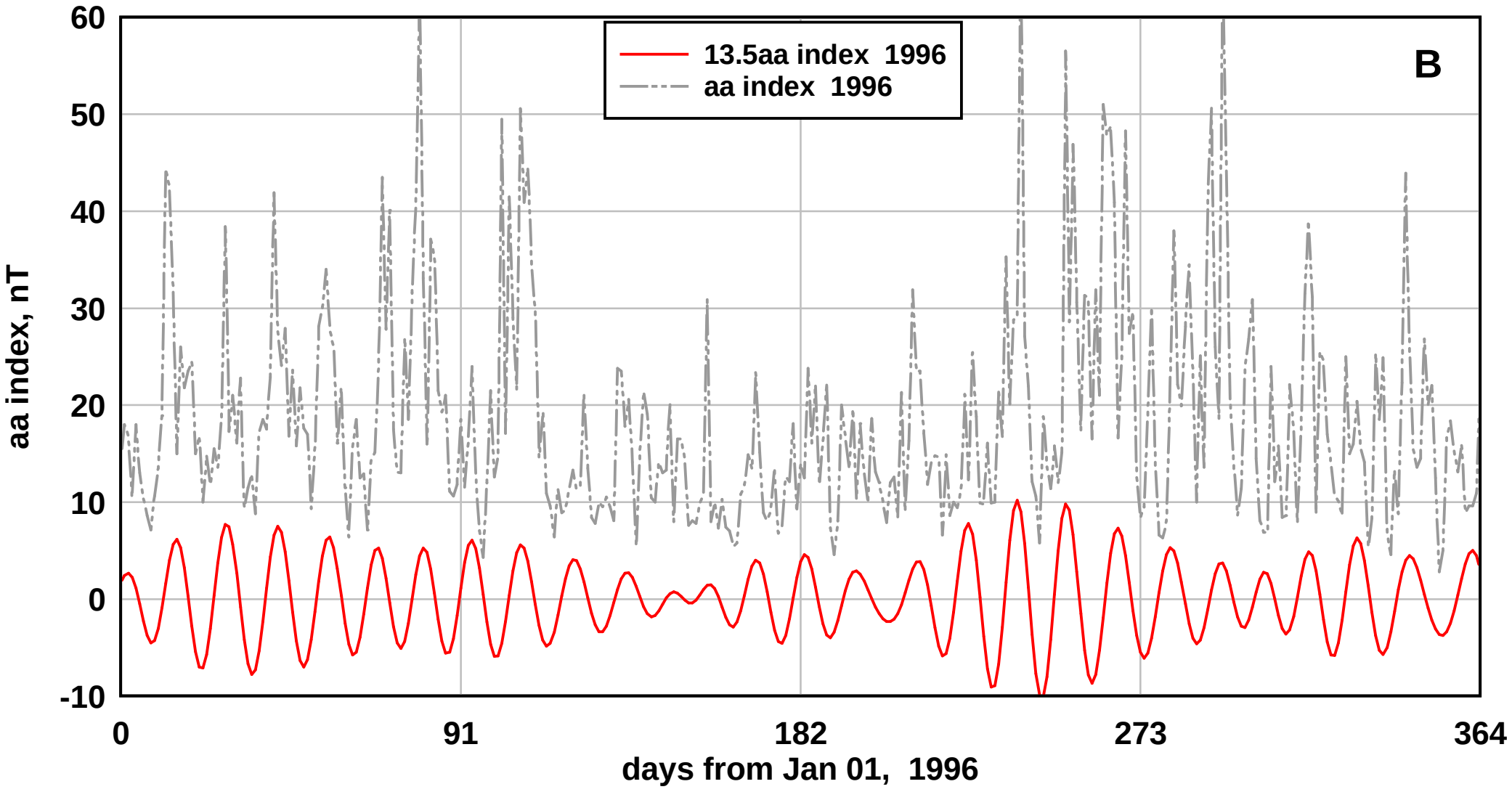

**Figure 1 A and B.** Compare aa index with the 13.5 day component of aa index for the years 1954 (A), and 1996 (B), years when the semi-annual variation was “ideal”, Cliver et al (2004).

Figures 1A and 1B show the aa index variations during 1954 and 1996. The time variation of aa index is usually highly variable from year to year. However, 1954 and 1996 were selected by Cliver et al (2004) as the two years out of the 130 years in the record that were closest to their measure of an "ideal" semi-annual variation. We note that in 1954, Figure 1A, and 1996, Figure 1B, there is also evidence of a quad-annual variation in the ~13.5 day component of the aa index. Figures 1C (1974) and 1D (2006) are examples of years when the semi-annual variation is much less evident while the quad-annual variation in the ~13.5 day component is clearly evident. Geomagnetic activity in the year 1974 has also been studied extensively as 1974 is exceptional in that all measures of solar and geomagnetic activity were relatively constant during 1974 e.g. Tsurutani et al (2006).

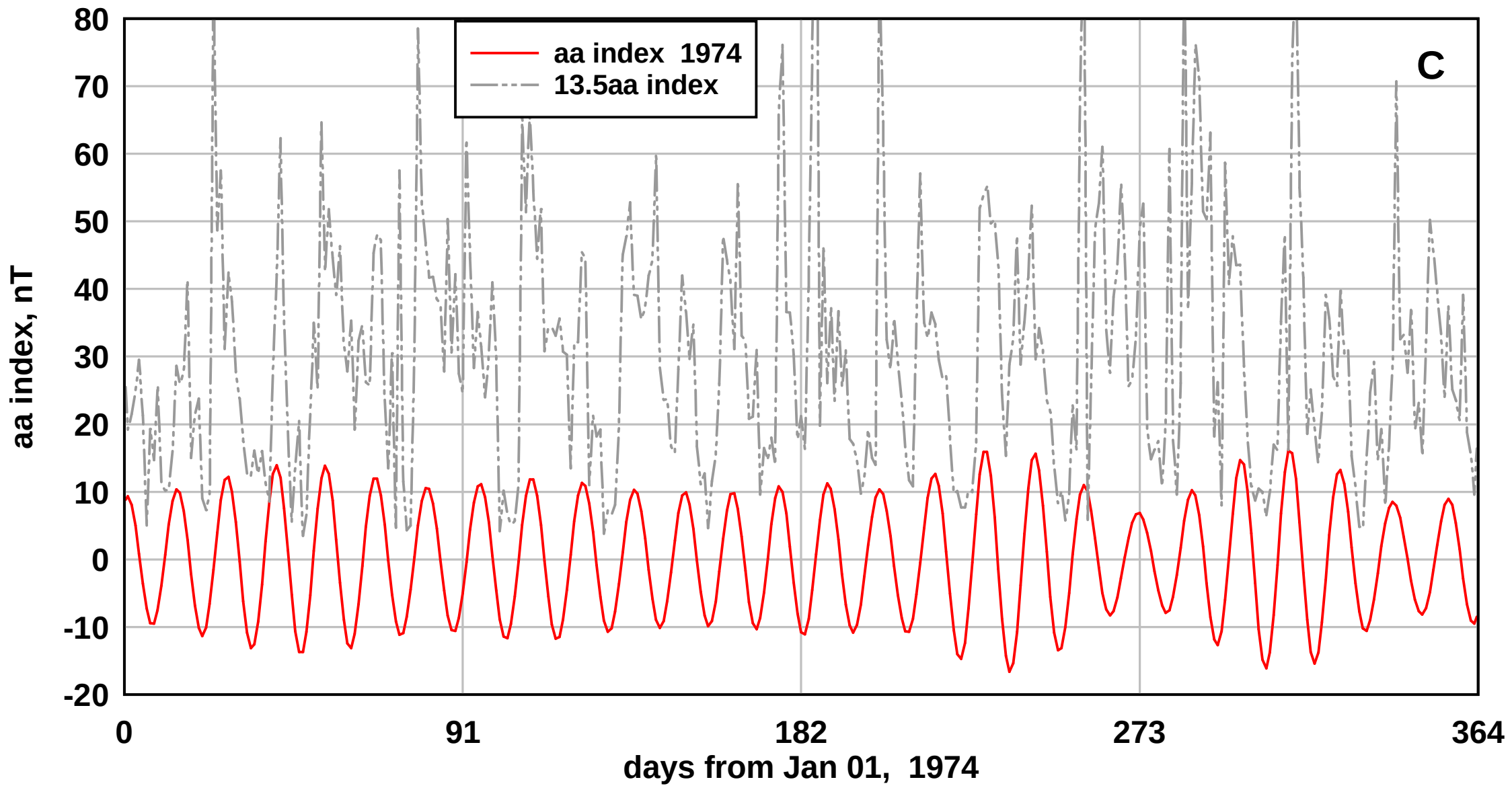

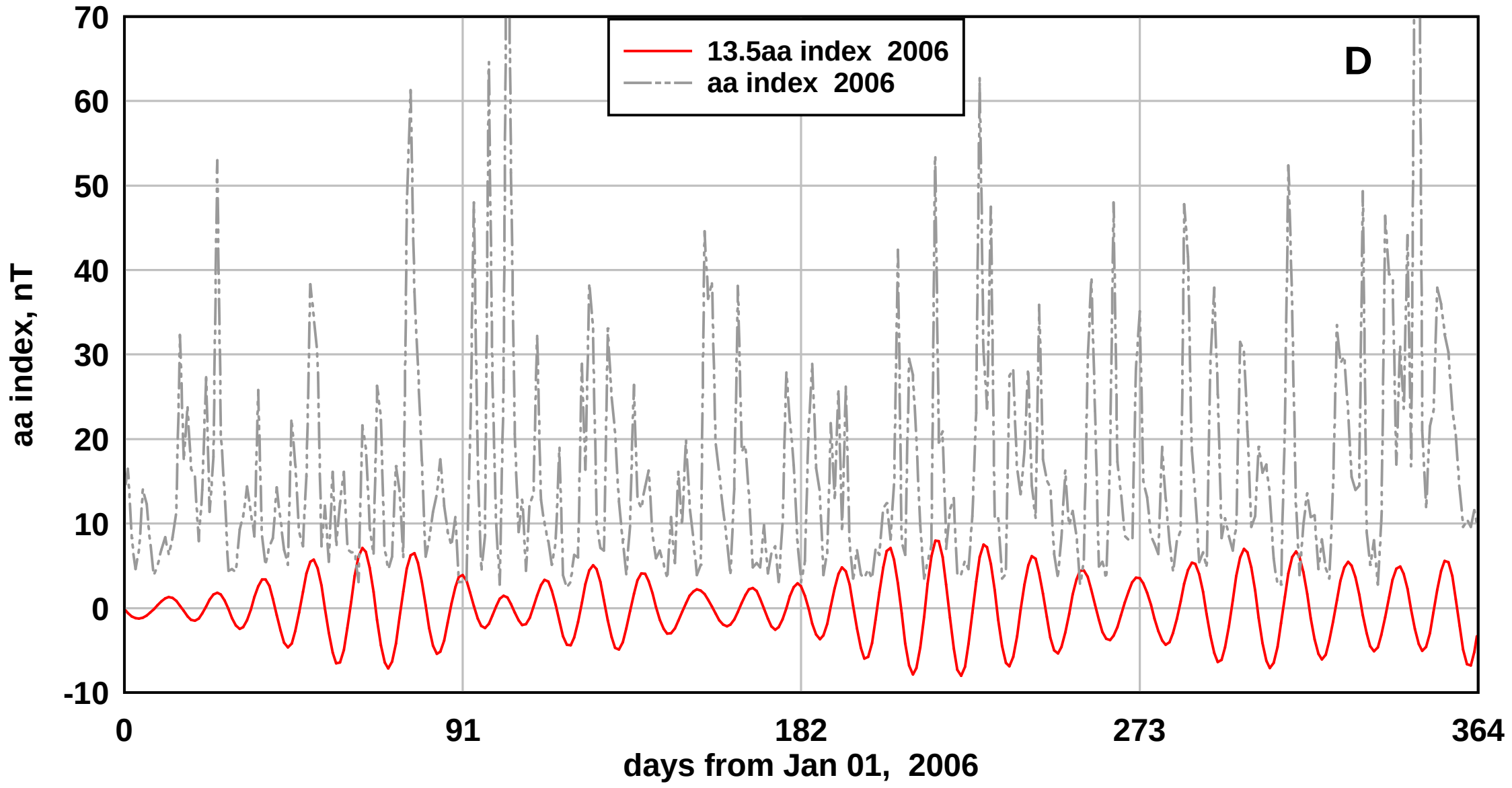

**Figures 1C and 1D.** Compare aa index with the 13.5 day component of aa index for the years 1974, (C) and 2006,(D).

A semi-annual variation in the ~13.5 day component of aa index is also often evident. Figures 2A and 2B show examples of years when the variation in the 13.5 day component would be regarded as primarily semi-annual. However, it is evident that a quad-annual variation is also present in each year. We will show later that the time variation of the ~13.5 day component of aa index varies systematically between predominantly semi-annual variation and predominantly quad-annual variation with elements of both types of variation usually present to some degree in any year. We will also show that the predominantly semi-annual variation occurs about three times more often and with higher intensity than the predominantly quad-annual variation.

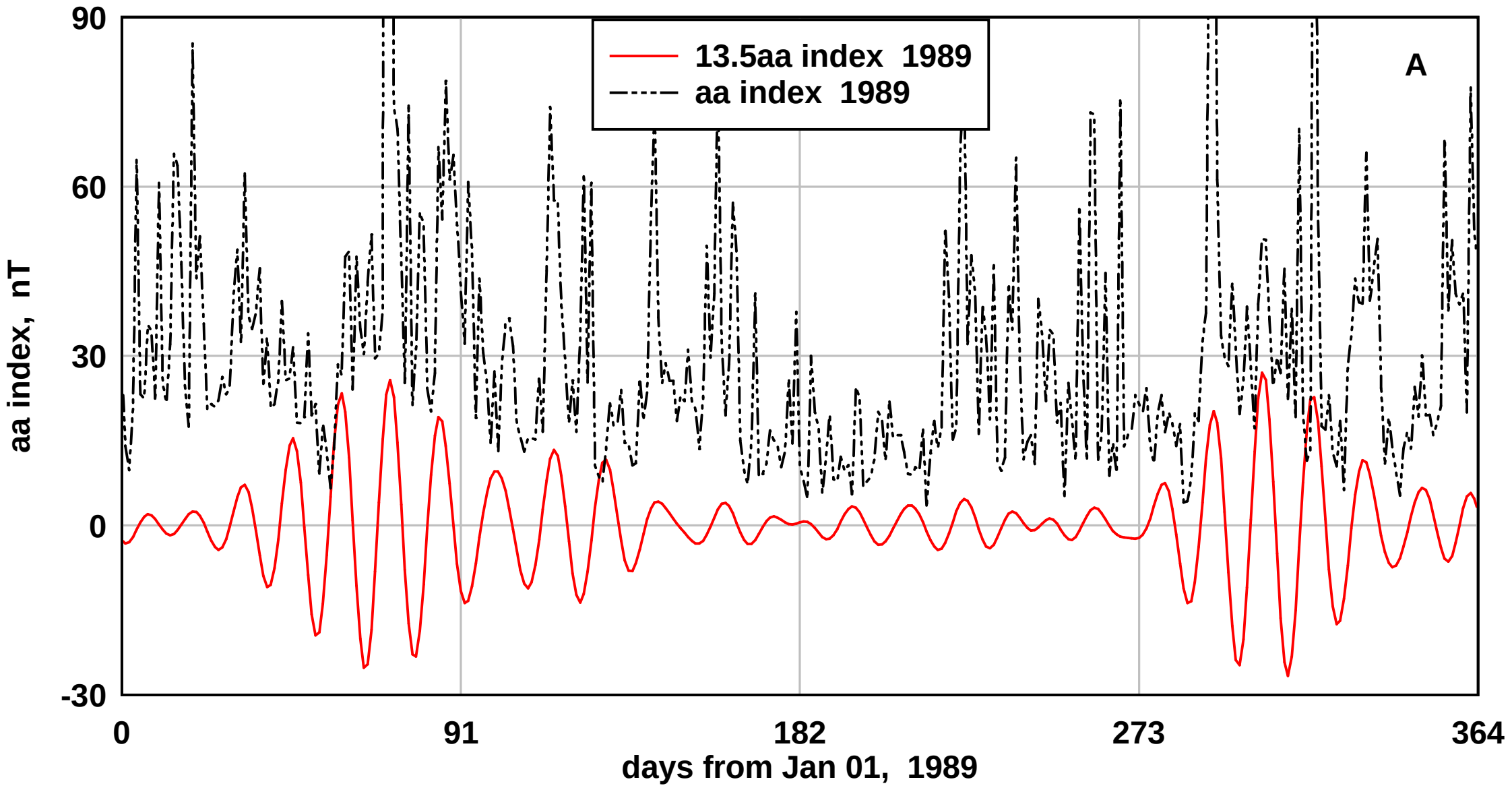

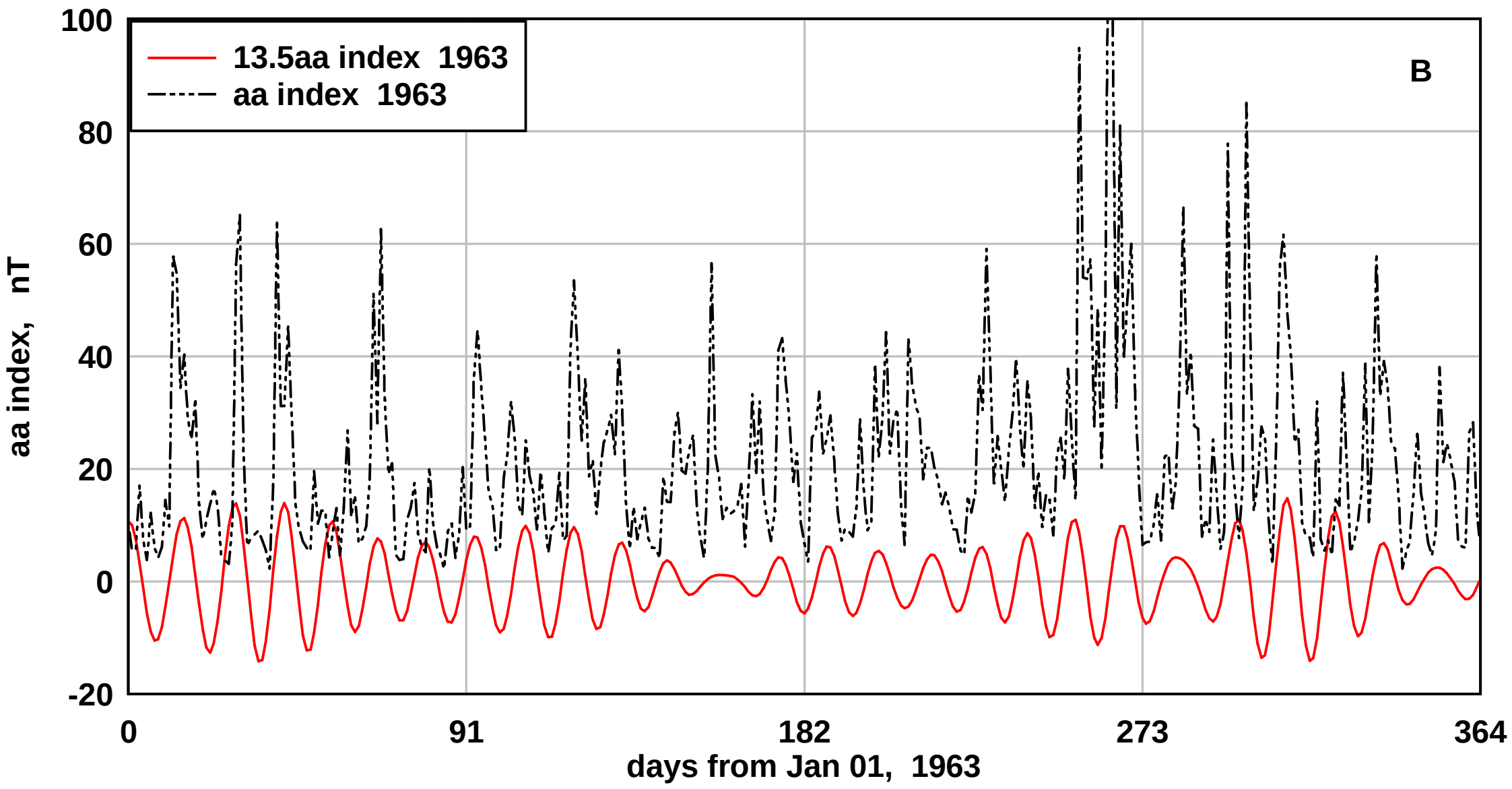

**Figures 2A and 2B.** Compare aa index with the ~13.5 day component of aa index for the years 1989, (A) and 1963, (B).

To the authors knowledge a quad-annual variation in geomagnetic activity has not been discussed in any of the numerous prior studies of geomagnetic activity. However, a number of prior studies provide evidence of a quad-annual variation superimposed on a larger semi-annual variation although the authors in each case do not discuss this aspect of their results. Examples are Cliver et al (2002), Figure 2, and Mc Pherron et al (2009), Figure 4, Russell and McPherron (1973), Figure 2.

In this paper we will attempt to explain the existence of a quad-annual variation in the ~13.5 day component of the aa index as well as explaining the systematic times of occurrence of predominantly semi-annual variation or predominantly quad-annual variation. Having unsuccessfully attempted to explain the observations described in Figures 1 and 2 using existing mechanisms that have been associated with the semi-annual variation we here consider the possibility that the observations in Figure 1 and 2 may be due to the influence of a planetary tidal effect on solar activity with the quad-annual variation arising from the tidal effect of Mercury on solar activity and the systematic change from predominantly semi-annual variation to predominantly quad-annual variation due to (a), the combined tidal effect of Mercury and Jupiter on solar activity and (b), the equinoctial effect on the transfer of heliospheric energy to the magnetosphere. This is a controversial approach as a planetary tidal effect on solar activity has, historically, been supported and rejected by numerous studies, Charbonneau (2002). A planetary tidal effect on solar activity is rejected, principally, as being too small to have any significance, e.g. de Jager and Versteegh (2005). However, there are other studies that provide support for such an effect e.g. Bigg (1967), Scafetta (2012) and references therein. Bigg (1967), using techniques from radiophysics for detecting periodic signals buried in noise showed that daily sunspot numbers for the years 1850 – 1960 contained a small but consistent periodicity at the 88 day period of Mercury, which is partially modulated by the position of Venus, Earth and Jupiter. Scafetta developed techniques for assessing the tidal effect of the eight significant planets, including the effect of planetary conjunctions, and the resultant effect on solar irradiance. The resulting time variation of the tidal effect is, with eight planets contributing, extremely complex. However, by taking 100 day averages Scafetta was able to isolate longer term cycles such as the 11 year solar cycle from the shorter term complexity. The reason, in this paper, for considering a planetary tidal effect is that the 88 day periodicity of Mercury is close to quad-annual, i.e. 365/88 = 4.1, and therefore offers the possibility of an explanation of the observations in Figures 1 and 2. In this paper we avoid the shorter term complexity of eightfold planetary conjunctions by considering only the tidal effects of Mercury and Jupiter. As shown below, the simple tidal effect model developed is successful in explaining much of the detail in the observed variation of the amplitude of the ~13.5 day component of the aa index.

The arrangement of the paper is as follows. Section 2 outlines the band pass filter method used to isolate the ~13.5 day component of aa index and specifies the data sources. Section 3 develops a simple model of the tidal effect on solar activity and outlines how the solar activity influences geomagnetic activity and the ~13.5 day component of aa index. Section 4 compares the predicted and observed variations of the amplitude of the ~13.5 day component of aa index. Section 5 discusses the results and Section 6 draws conclusions.

## 2. Data analysis.

The data used in this study are records of geomagnetic aa index (aa) and solar wind velocity, (SWV). Isolation of the ~ 13.5 day component of each variable was obtained by making a Fast Fourier Transform of each data series. The resulting n Fourier amplitude and phase pairs, $A_n(f_n)$, $\phi_n(f_n)$, in the frequency range 0.062 days$^{-1}$ to 0.090 days$^{-1}$ (period range 16 to 11 days) were then used to synthesize a band pass filtered version of each variable, denoted for example 13.5aa, by summing the n terms, $aa_n = A_n\mathrm{Cos}(2\pi f_n t - \phi_n)$ for each day in the series. Where a data series has been smoothed by, for example, a 30 day running average, the resulting smoothed series is denoted e.g. aa S30.

## 3. Development of a model of a tidal effect on solar activity and geomagnetic activity.

A quad-annual effect repeats four times per year or once every 91 days. As mentioned above failure to find an explanation of the quad-annual variation of aa index from the earlier mechanisms mentioned in the introduction led to examining if the variation might be associated with the periodicity of Mercury, $T_M$ = 88 days, and/or the periodicity of the conjunction of Mercury and Jupiter, $T_{MJ}$ = 90 days. The period of conjunction of two planets is $T_{12} = (1/T_1 - 1/T_2)^{-1}$. The 10 point smoothed frequency spectrum of the daily variation of aa index, 1868 to 2012, is shown in Figure 3. With 52,960 data points the resolution in the spectrum, even at low frequency, is high. The semi-annual peak at 182.6 days is marked SA and the next highest peak, at the period of Mercury, 88 days, is marked $T_M$.

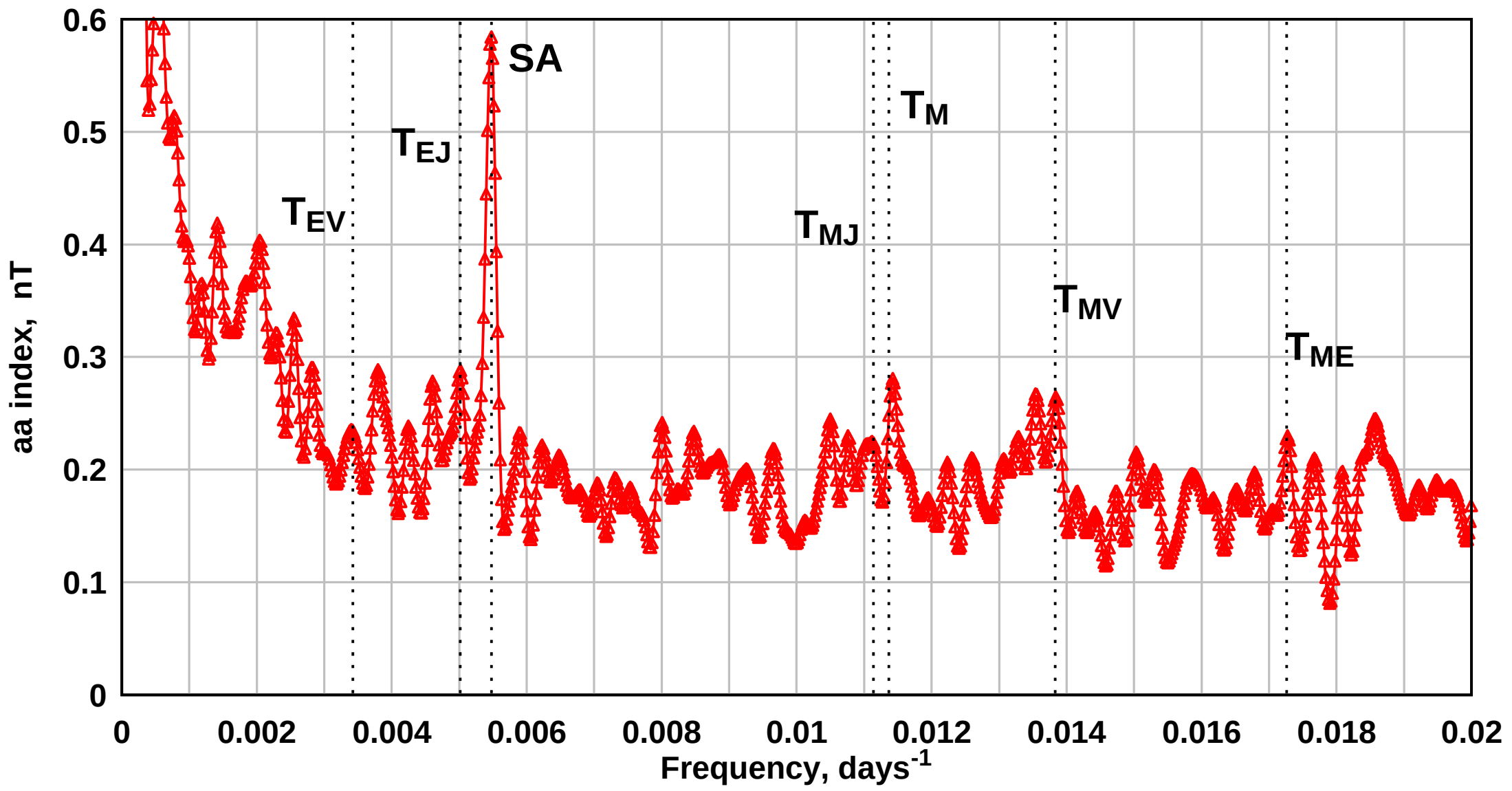

**Figure 3.** The frequency spectrum of the daily aa index 1868 to 2012. The semi-annual variation (SA) and the orbital period of Mercury ($T_M$) are indicated as well as some possible planetary conjunction periods.

There is no evidence of a peak at the second harmonic of the semi-annual variation so we discount the possibility that the quad-annual observations in Figures 1 and 2 could arise from the second harmonic of the semi-annual variation. Also marked are the periods of various planetary conjunctions. In particular the period, $T_{MJ}$ = 90 days, of the Mercury-Jupiter conjunction is, as expected, close to the orbital period of Mercury. With some effort, outside the scope of this paper, many of the peaks in Figure 3 can be linked to other orbital periods and planetary conjunction periods. For example the peak at 0.0084 days$^{-1}$ may be due to the Venus-Jupiter conjunction. Triple near conjunctions may also feature and methods for calculating the periods of multiple conjunctions can be found in Hung (2007) and Scafetta (2012). The amplitude of the $T_M$ peak is about 0.1 nT, about 25% of the amplitude of the SA peak, 0.4 nT. This suggests that a modulation of solar activity at the period $T_M$ could contribute to the variation of the aa index at about one quarter of the level of the semi-annual variation.

We calculate the tidal effect of a planet from the ratio $Mp/Rp^3$ where Mp is the mass of the planet in kg and Rp is the distance between the planet and the Sun in A.U. Rp in A.U. is found for the various planets as a function of day of year at http://cohoweb.gsfc.nasa.gov/helios/plan_des.html . The masses of the planets, Jupiter, Mercury and Earth are respectively, $M_J$ = 1.90E27, $M_M$ = 3.30E23 and $M_E$ = 5.97E24 kg. The daily variation of the tidal effects of Jupiter, Earth and Mercury are shown for the years 1995 to 2005 in Figure 4A. The sum of the tidal effects of Mercury and Jupiter during the eleven year period 1995 to 2005 is shown by the full line, (y1 + y2), in Figure 4A. We ignore the tidal effect of planets other than Mercury and Jupiter in this study.

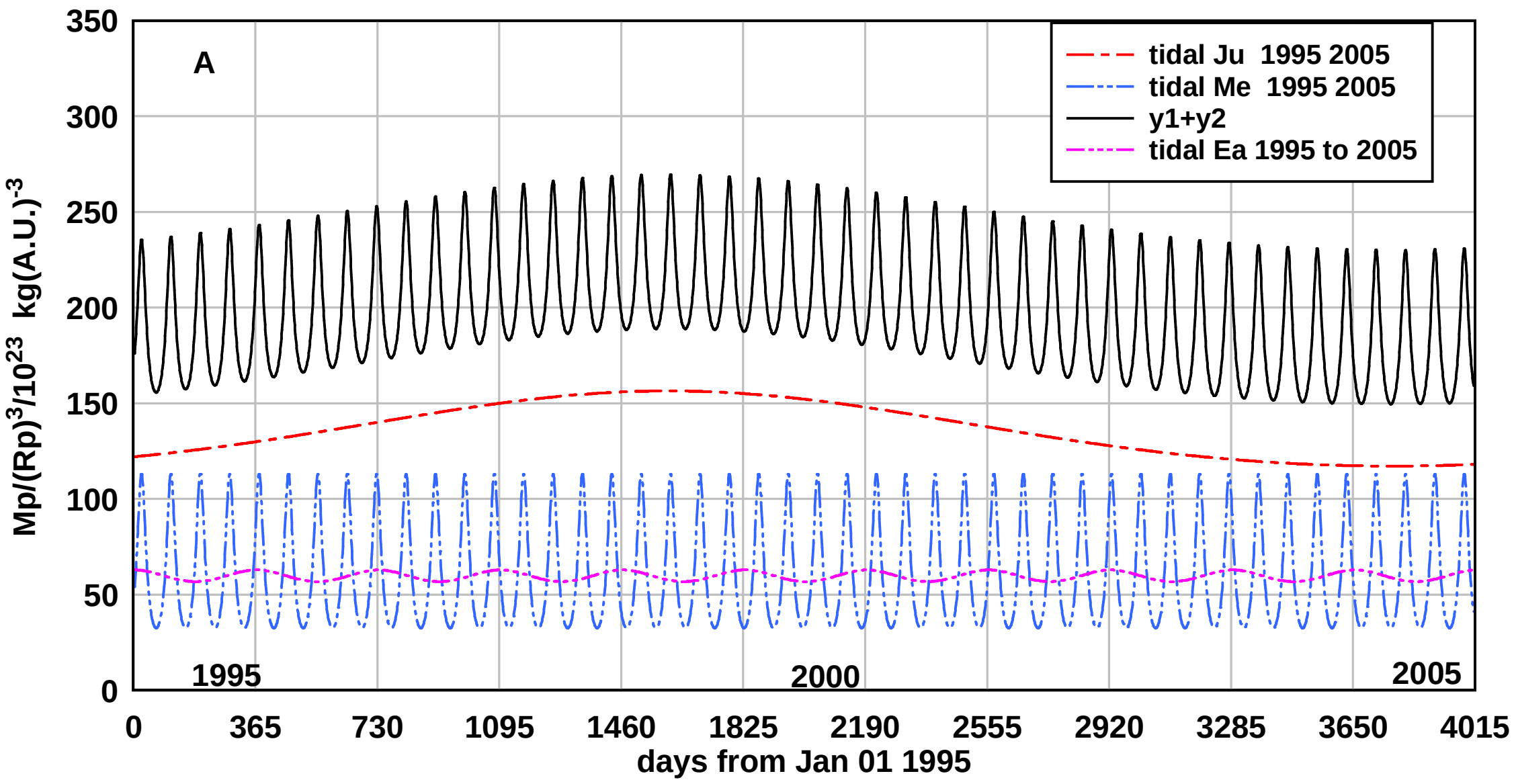

**Figure 4A.** The tidal effect of Mercury, Jupiter and Earth at the Sun for years 1995 to 2000.

The model proposed here is that the combined tidal effect has an effect on solar activity when some threshold level of the tidal effect is exceeded. Arbitrarily we choose the level 200E23 kg(A.U.)$^{-3}$ in Figure 4A as that threshold. The resultant tidal effect i.e. the tidal effect above 200 units is shown in Figure 4B and denoted “tidal effect > 200”. We assume that solar activity is proportional to the tidal effect above the threshold so that solar activity is proportional to the level of the curve labeled “tidal effect > 200” shown in Figure 4B.

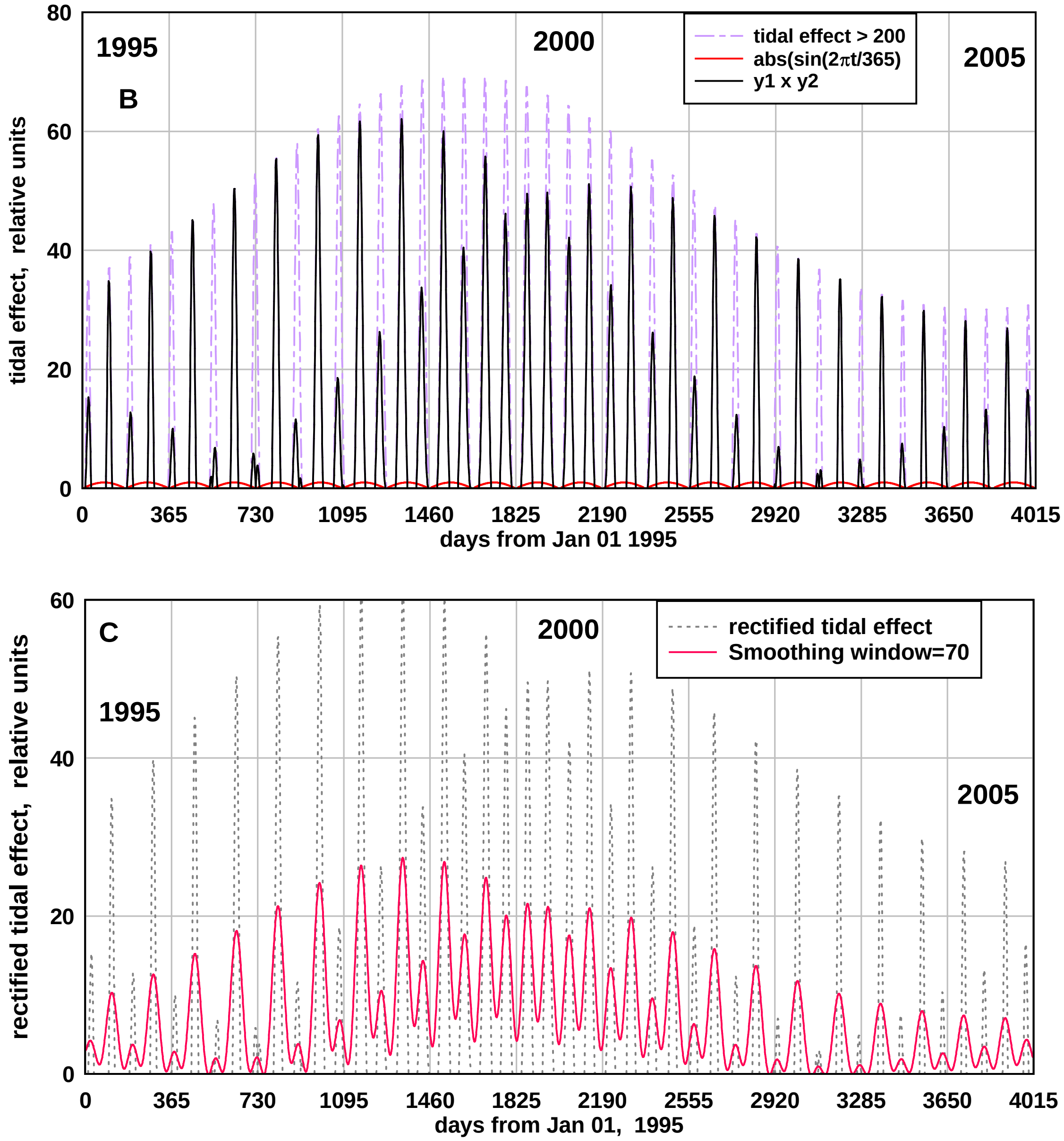

**Figure 4B and 4C.** (B), the tidal effect of Mercury and Jupiter > a threshold of 200 units, the equinoctial rectifier function, abs(sin(2πt/365)), and the product of the tidal effect, y1, and the rectifier function, y2. for the years 1995 to 2000. (C), the 70 day running average of the rectified tidal effect.

We next assume the concept of the magnetosphere acting as a rectifier to modulate the solar activity from the Sun so that more energy is injected into the magnetosphere during equinoctial months than during solstitial months, Russell and Mc Pherron (1973). For simplicity we assume the rectifier effect is given by abs(sin(2πt/365)). Thus the rectified tidal effect is the product of the tidal effect and the rectifier effect: [tidal effect (Ju + Me) > 200][abs(sin(2πt/365))]. The rectified tidal effect is shown in Figure 4B. Clearly the effect in most years is a predominantly semi-annual effect having two major maximums around equinox. However, a quad-annual effect, i.e. four approximately equal peaks in activity each year, is seen to emerge for one or two years around year 2000. These effects are seen more clearly when the data in Figure 4B is smoothed by a 70 day running average, Figure 4C. Every seven years the phase of the tidal effect of Mercury and Jupiter returns to the same annual phase relationship. This occurs because 7x 365.256/87.96926 = 29.07, very close to a whole number. Thus if a predominantly quad-annual effect occurs in year 2000 another interval of quad-annual effect can be expected seven years later in year 2007 or seven years earlier in 1993.

Two rather arbitrary assumptions were made in deriving this model. First, 200E23 kg(A.U.)$^{-3}$ was chosen as the level above which solar activity responds to the tidal effect. The effect of varying this level by 10% to 220E23 units decreases the strength of the tidal effect and increases the proportion of quad-annual variation in each year. A second assumption, made in the interests of simplicity, is that the rectifier effect of the magnetosphere is simply abs(sin(2πt/365)). This infers that the response of the magnetosphere is zero at solstices and unity at equinoxes. A less severe rectifier effect is [1 + abs(sin(2πt/365))]/2 giving a response of 0.5 at solstices and 1 at equinoxes. This increases the strength of the tidal effect and, again, increases the proportion of quad-annual variation in each year. Figure 5 shows the result of the two changes and comparing Figure 5 with Figure 4C illustrates that the changes result in a shift towards the occurrence of quad-annual variation rather than a semi-annual variation. However, the timing of the occurrence of predominantly quad-annual effect remains the same. It may be noted that if the arbitrary threshold at 200 units were increased by 25% to 250 units a tidal effect would not exist for more than half the time and if increased, a little more, to 270 units a tidal effect would not exist at any time.

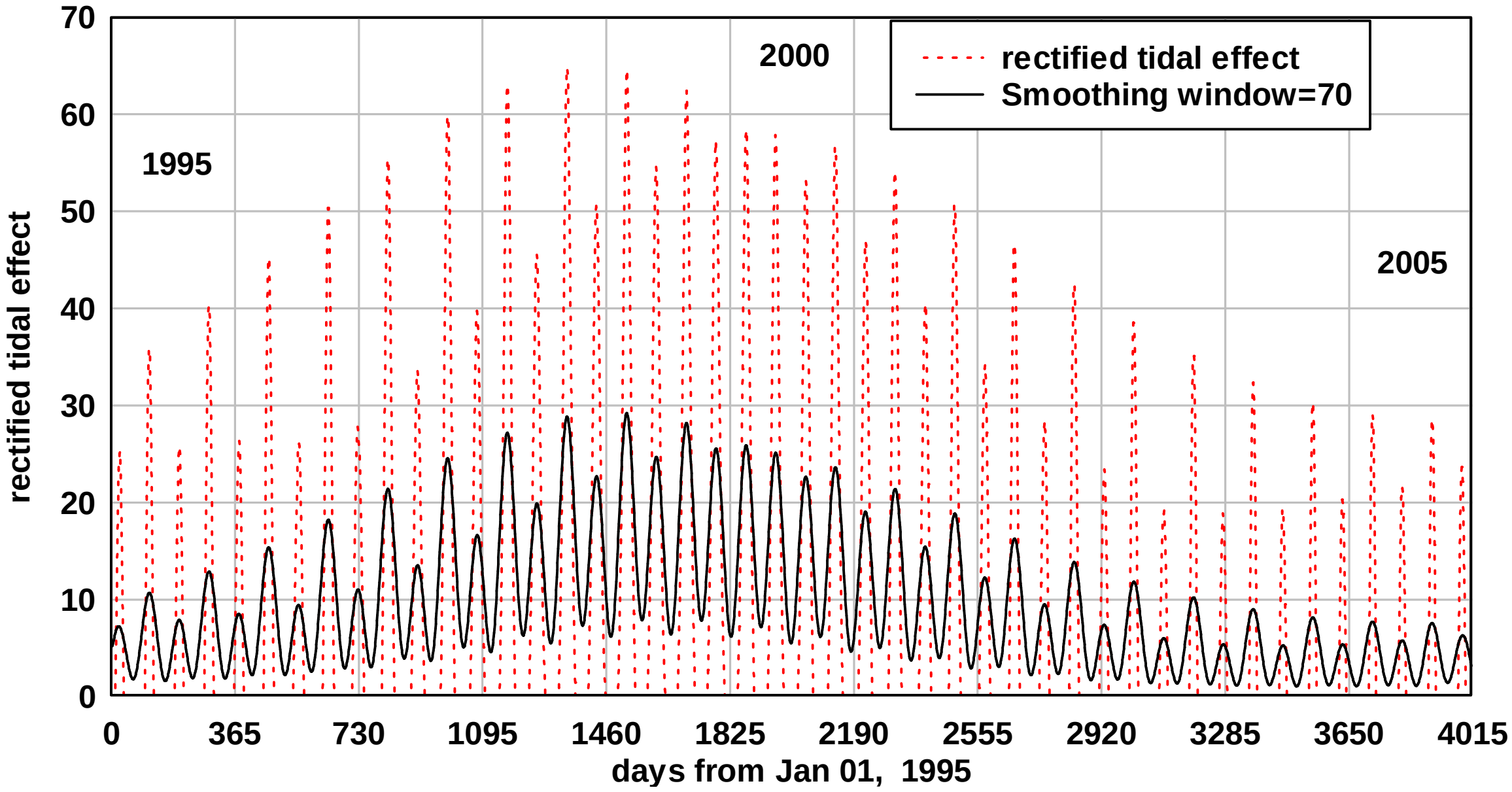

**Figure 5.** The rectified tidal effect when the threshold is set at 220 tidal units and the rectifier effect response is the less severe [1 + abs(sin(2πt/365))]/2.

In summary the tidal model predicts a seven year cycle between events of predominantly quad-annual variation in geomagnetic activity. Before moving to the next section where the predicted variation in aa index is compared with the observed variation we make some comments comparing the model with observations from previous studies.

(1) If solar activity is influenced by a planetary tidal effect then one intuitively expects that the solar activity would occur, like the tidal elongation, on opposite sides of the Sun. Again, intuitively, one expects that solar activity occurring on opposite sides of the Sun would lead to a 13.5 day variation in terrestrial effects. Thus the quad-annual effect predicted by the model would be expected to be most noticeable in the ~13.5 day component of the aa index and less noticeable in the ~27 day or ~9 day components.

(2) The model explains the predominance of semi-annual or equinoctial maximums in long term averages of geomagnetic activity because, as shown above, a predominantly semi-annual variation occurs in about five of every seven years. However, in this model the maximums, either predominantly semi-annual or predominantly quad-annual or some combination, are due to modulation of solar activity by the combined tidal effects at Mercury's 88 day periodicity and Jupiter's 12 year periodicity. In prior mechanisms relating to the semi-annual effect it is inferred that, on an annual basis, solar activity is essentially constant or randomly distributed and the semi-annual variation in geomagnetism is a result of selection of the solar activity by an axial, equinoctial or Russell-McPherron effect.

(3) The model explains the variability of the dates of the maximums in average semi-annual variation reported by Chapman and Bartels (1940) and commented on by Russell and McPherron (1973). For example the model predicts that in 1998 (days 1095 – 1460 in Figure 4C) the major maximums occur earlier in the year than equinox while in 2002 (days 2555 to 2920 in Figure 4C) the major maximums occur later in the year than equinox.

(4) As the model favors a semi-annual variation about 5/7 of the time the model explains why most magnetic storms occur around equinox including the historically large magnetic storms.

(5) The model explains why averages of the variation of geomagnetic indices or relativistic electron flux over a solar cycle usually exhibit two peaks close to the equinoxes, Baker et al (1999) and Li et al (2001). The reason for each equinox having two peaks in a cycle average is evident from Figure 4C. During the ascending phase of solar cycle 23, (years before 2000 in Figure 4C), the major peaks occur a little earlier than equinox while during the descending phase of cycle 23, (years after 2000 in Figure 4C), the major peaks occur a little later than equinox. Therefore a cycle average would tend to show two peaks on either side of the equinoxes.

(6) The model explains why activation of ~27 day and ~13.5 day components in solar wind velocity, temperature, ion density and Kp index typically last for two to three solar rotations i.e. for 54 to 81 days, Mursula and Zeiger (1996). The model is based on a tidal effect with periodicity of 88 days. The model therefore predicts activations lasting less than 88 days – consistent with two to three solar rotations.

## 4. Predicted and observed variation of the ~13.5 day component of aa index

The NASA Cohoweb site provides daily values of planetary radii for 1959 to 2019 that can be used as input to the tidal effect term, $(M_M/R_M^3 + M_J/R_J^3)$, in the model. Figure 6 shows the rectified tidal effect, [tidal effect (Ju+Me) > 200][abs(sin(2πt/365))], for 1959 to 2013. Due to the scale of the diagram the years with predominantly quad-annual variation show up as the more dense regions in the graph at intervals that alternate between 6 and 8 years with an average, for reasons discussed above of 7 years. There are 9 intervals of predominantly quad-annual variation in Figure 6. Also indicated in Figure 6, by the symbol mx, are the approximate years of solar cycle maxima. It is evident that the five broad maxima in the tidal effect do not coincide with solar cycle maxima. The average period of tidal effect maxima over this interval is about 11.8 years while the average solar cycle period is about 10.6 years

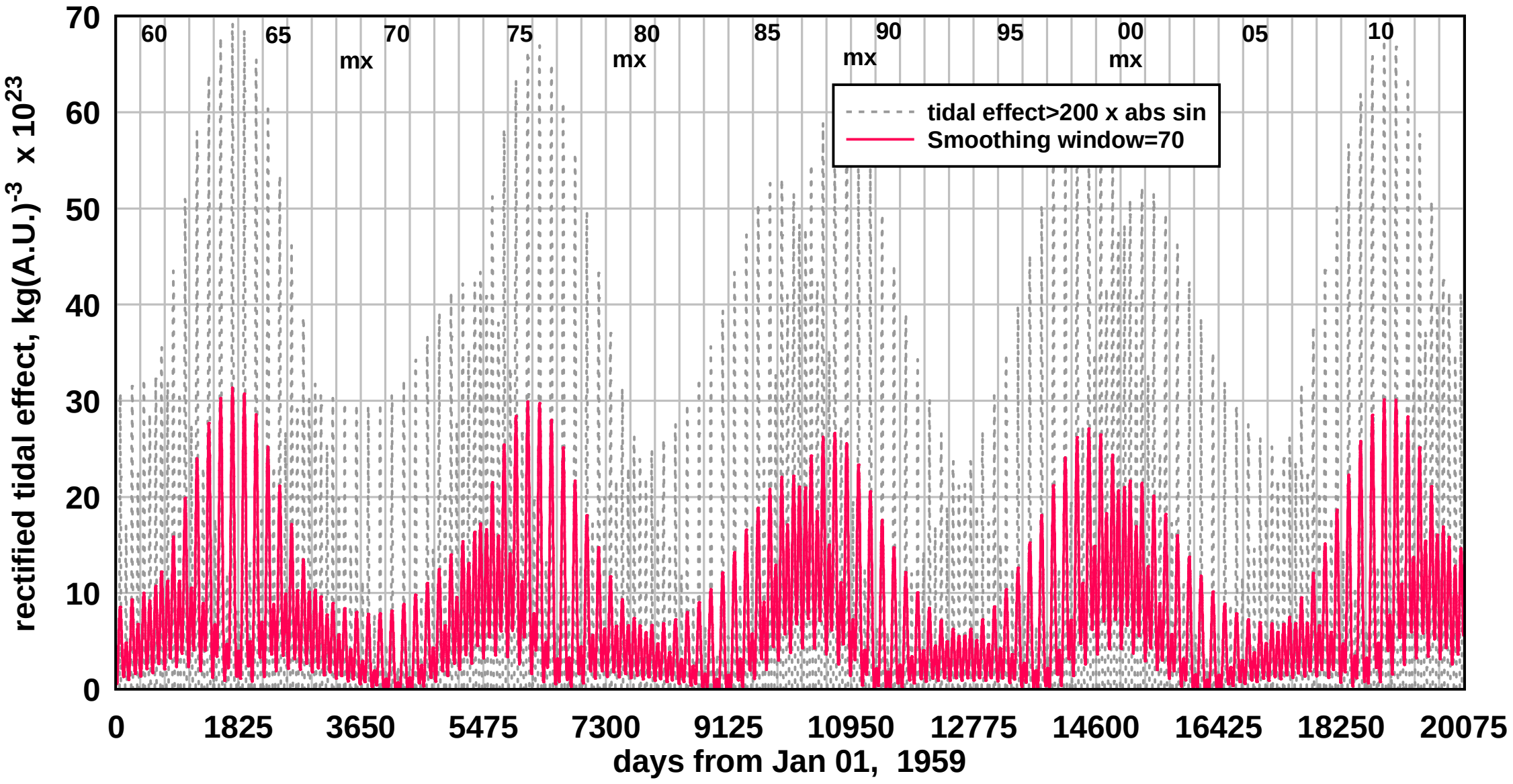

**Figure 6.** The rectified tidal effect for the years 1959 - 2013.

Figure 7A - 7C and Figure 8A – 8C are expanded versions of Figure 6 with the 20 day smoothed amplitude of the ~13.5 day component of the aa index superimposed for comparison. Here we are comparing a rather complex but systematic tidal effect variation with a signal, the ~13.5 day component of aa index, that has a large noise component. We intuitively expect that the noise in the aa index variation will be greater around times near solar cycle maximum and therefore expect correlations between the predicted tidal effect variation and the observed aa index variation to be more evident at years near solar cycle minimum.

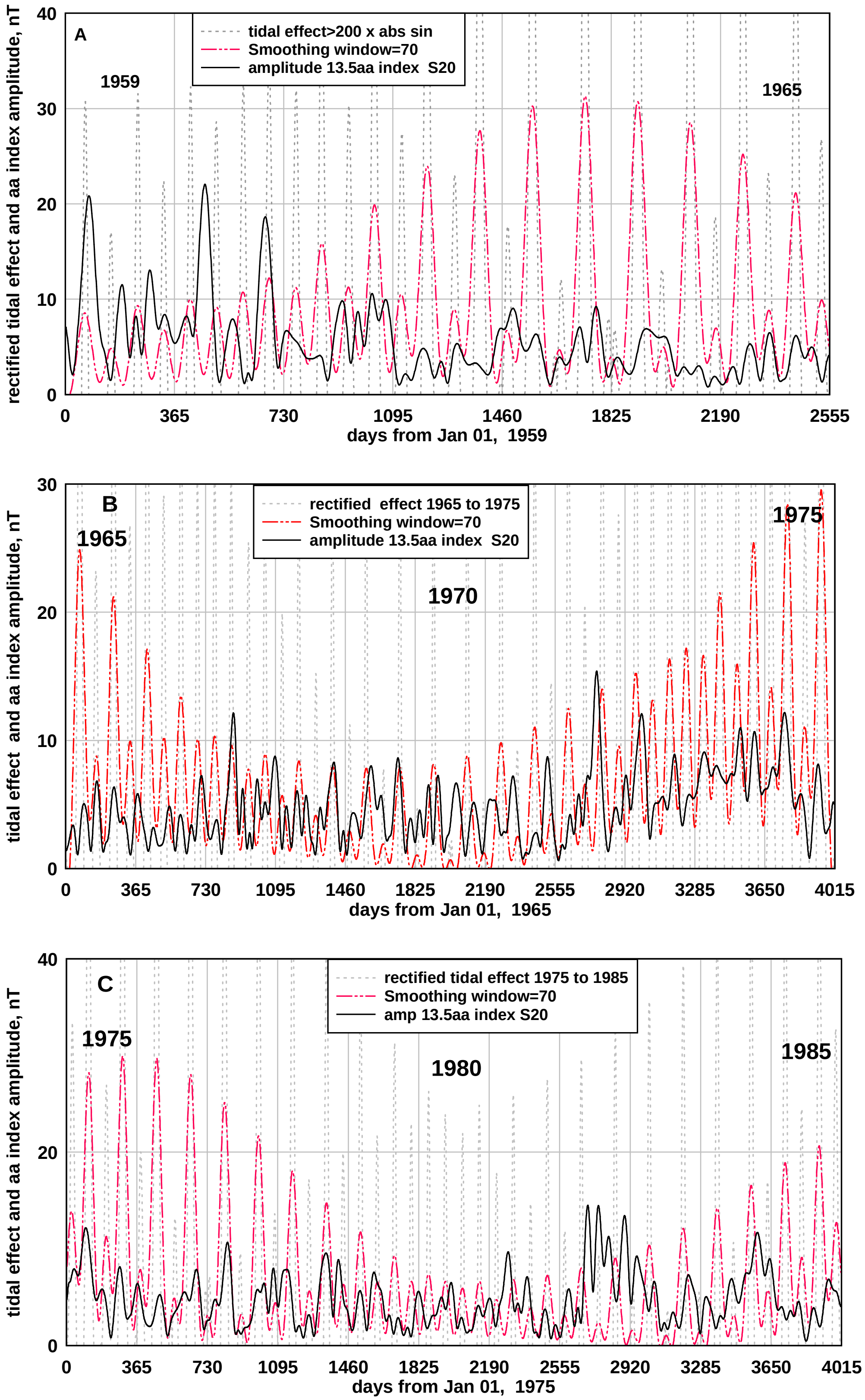

**Figure 7.** Comparison of the rectified tidal effect with the amplitude of the 13.5 day component of aa index. (A) 1959 – 1965, (B) 1965 - 1975, (C) 1975 - 1985.

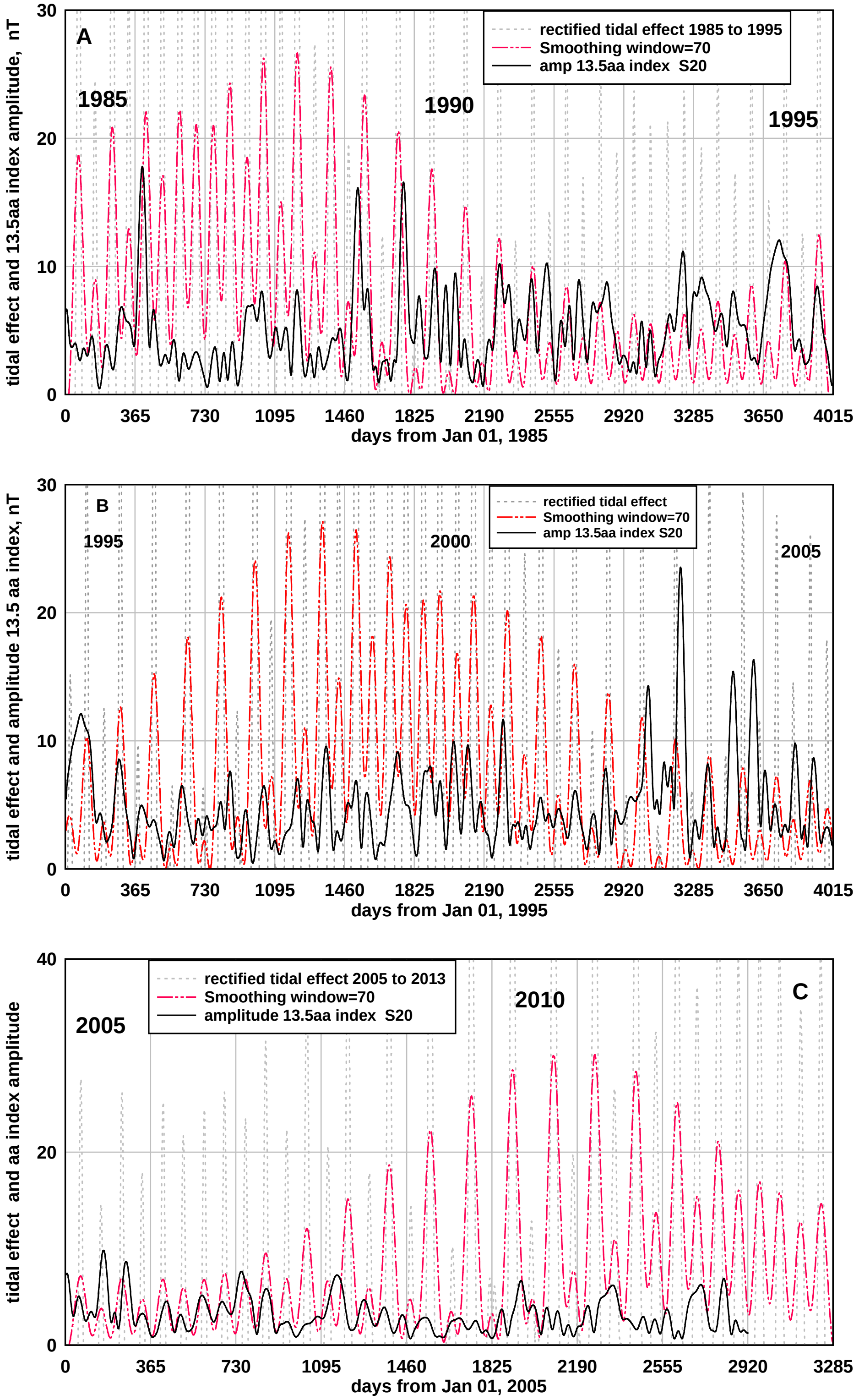

**Figure 8.** Comparison of the rectified tidal effect with the amplitude of the 13.5 day component of aa index. (A) 1985 - 1995, (B) 1995 - 2005, (C) 2005 - 2013.

Some correlations are immediately evident for the example years discussed in the introduction. The year 1996 was one of the two years selected by Cliver et al (2004) as evidencing their "ideal" of semi-annual variation. Figure 6 and Figure 8B indicate that the model predicts that the tidal effect variation in 1996 will be predominantly semi-annual with the maxima very close to the equinoxes i.e. a highly symmetrical semi-annual variation consistent with the Cliver et al (2004) finding that the variation in 1996 in an "ideal" semi-annual variation. The variation in 1996 can be compared with the variation in 2001 where the model predicts the occurrence of two maxima each at times significantly later than equinox and the observed aa index variation corresponds closely to this prediction.

Working through the sequence of observations in Figure 1 and Figure 2 we note similar correlation of observations with prediction. In 1974 the predicted variation (Figure 7B) is for a predominantly quad-annual variation superimposed on a significant constant level and this coincides fairly closely with the observed variation of aa index amplitude in Figure 7B and in Figure 1C. In 2006 the prediction (Figure 8C) is for a predominantly quad-annual variation superimposed on a relatively low constant level again coinciding very closely with the observed variation of aa index amplitude in Figure 8C and Figure 1D. In 1989 the prediction (Figure 8A) is for a predominant and strong semi-annual variation superimposed on a relatively low constant level coinciding very closely with the observed aa index variation in Figure 8A and Figure 2A. In 1963 the prediction (Figure 7A) is for a predominantly semi-annual variation superimposed on a relatively low constant level coinciding very closely with the observed aa index amplitude variation in Figure 7A and Figure 2B. There are numerous other years of very good correlation between prediction and observation e.g. 2000 (Figure 8B), 1977 (Figure 7C), 1969 (Figure 7B), 1975 (Figure 7B), 2005 (Figure 8B). In fact for about half of the 50 years in the sample the correlation ranges between reasonable and very good. We note reasonable correlation with the quad-annual variation in years 1960, 1965, 1966, 1974, 1975, 1978, 1979, 2000, 2002, 2005, 2006 and 2007. In view of the very noisy nature of the aa index the observed correlations provide convincing support for this very basic model of planetary tidal effect on solar and geomagnetic activity. Given that there is a planetary tidal effect then one might expect that other planets besides Mercury and Jupiter would be included in a more detailed model. For example a more detailed model might include, as well as modulation by the tidal effect of Mercury and Jupiter, modulation by the tidal effect of Earth, Venus and Saturn and modulations at the periods of conjunction between the aforementioned planets, Bigg (1967), Hung (2007) and Scafetta (2012).

## 5. Matters arising from the results.

### 5.1 Origins of the quad-annual variation in geomagnetic activity

The basis of the model and the results in the previous section is that the semi-annual and quad-annual variations in the ~13.5 day component of geomagnetic activity arise primarily from an 88 day period variation in the strength of a planetary tide on the Sun and the resulting 88 day period variation in solar activity. Following the hypothetical line of connection, tidal variation in solar activity, presumably in the form of increased coronal holes and open solar flux, leads to 88 day period variation in solar wind speed and interplanetary magnetic field. The interaction of the solar wind with the magnetosphere results in a semi-annual or quad-annual variation in the amplitude of geomagnetic activity and in the amplitude of the aa index. Thus, the observation of an 88 day period component in solar wind velocity would provide support for the model and the connection outlined. As variations in the solar wind velocity and in the aa index are, similarly, very noisy we adopt the approach taken by Cliver et al (2004) in their investigation of the origins of the semi-annual variation in geomagnetic activity of selecting for study years that show pronounced variation of the type under study. Here we are interested in examining if the origin of the quad-annual variation in geomagnetic activity has, as proposed in the model, originated from an 88 day period variation in solar activity and solar wind velocity. As we wish to detect the presence of an 88 day period in the solar wind velocity it is necessary to select sequences of observations long enough to have some prospect of resolving a component of this periodicity. The model predicts that quad-annual variation will be predominant for one or possibly two years and then systematically change to mainly semi-annual variation. The model calculation in Figure 6 shows that the model predicts nine short, ~ two year long, intervals of predominantly quad-annual variation spaced at intervals of about seven years during the 53 years between 1959 and 2010. To improve the frequency resolution in resultant spectra we select sequences that are three years long and, to reduce the effects of noise, close to solar cycle minimum. Figure 9A shows daily sequences of the ~13.5 day component of solar wind velocity for the years 1973 – 1975 and 2005 – 2006, the first sequence is shifted by +200 m/s for clarity. Figure 9B shows the frequency spectrum of the amplitude of the ~13.5 day solar wind velocity component in each sequence. Despite the low frequency resolution it is evident that there is a significant component of solar wind velocity close to the 88 day period of Mercury in each sequence. This observation supports the model of a planetary tidal effect modulating solar activity, solar wind velocity and leading to a quad-annual variation in geomagnetic activity during the selected years.

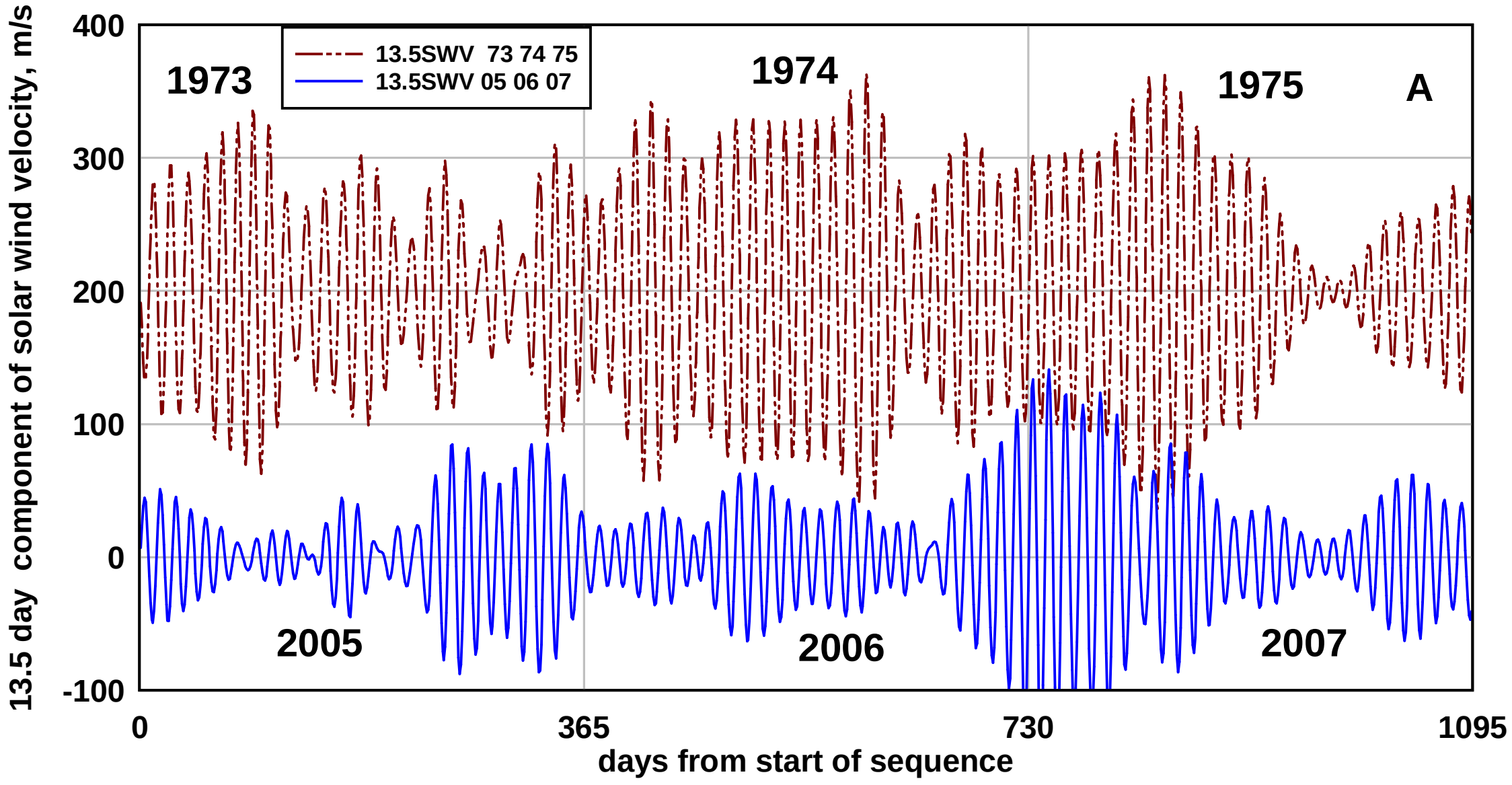

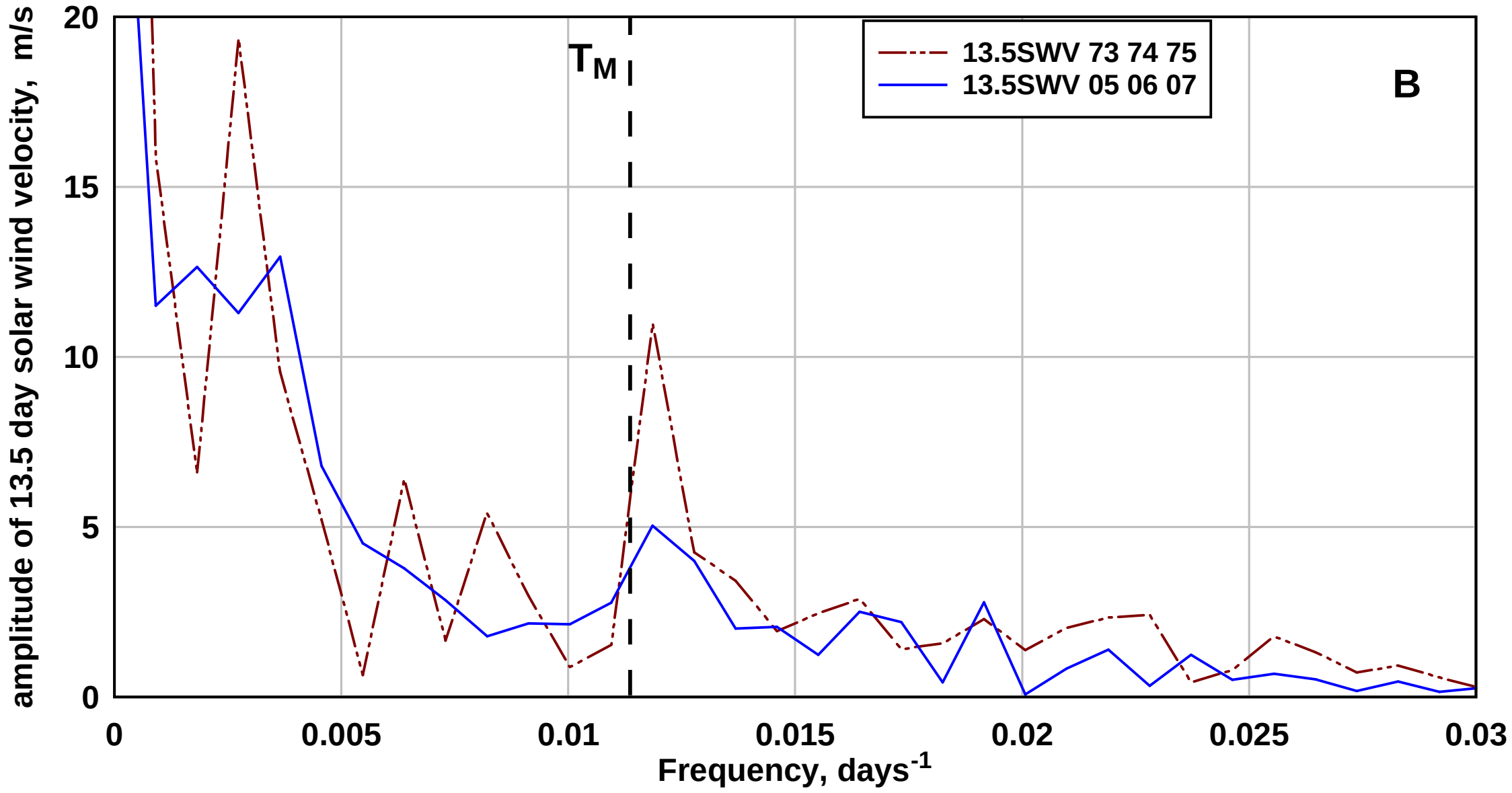

**Figure 9.** (A) The variation of the 13.5 day component of solar wind velocity 1973 – 1975 and 2005 – 2007. (B) The frequency spectra of the amplitude of the 13.5 day variations in (A).

**5.2 Origin of frequency components in the spectra of geomagnetic activity.**

In this section we extend the spectral range of the spectrum of aa index in Figure 3 from 0.02 days$^{-1}$ to 0.1 days$^{-1}$ to include the major solar rotation components around 27 day period and 13.5 day period, Figure 10. The dotted reference lines at the low frequency end of Figure 10 are at the same periods as the reference lines in Figure 3. Here we are interested in the origin of the two sidebands at each side of the 27 day peak that are evident at 25 and 29 days and the origin of the twin peaks close to 0.026 days$^{-1}$ at periods 37.4 and 38.7 days. The model outlined in Section 3 is based on an 88 day period amplitude modulation of solar activity, solar wind velocity and ultimately geomagnetic activity. As the series of daily average aa index is 142 years long it is expected that the 88 day period modulation of the solar rotation components and the resulting sideband peaks should be resolved in the frequency spectrum. Sidebands should occur at $f_1$ +/- $f_M$ and $f_2$ +/- $f_M$ where $f_1$ = 1/27, $f_2$ = 1/13.5 and $f_M$ = 1/88 days$^{-1}$. Thus the two first harmonic sidebands are expected at 0.0257 days$^{-1}$ (38.9 days) and 0.0484 days$^{-1}$ (20.7 days) and the two second harmonic sidebands are expected at 0.0627 days$^{-1}$ (15.9 days) and 0.0854 days$^{-1}$ (11.7 days). We also expect sidebands due to the equinoctial modulation $f_E$ = 1/182 days, and due to an annual modulation, $f_A$ = 1/365 days. However, it should be noted that a single, low frequency peak at $f_A$ is not evident in the observed spectrums, Figure 3 and Figure 10. Most of the aforementioned sidebands would be subject to modulation and fine splitting due to the 11 year solar cycle, the peak amplitude of which is ~ 1 nT, an order of magnitude higher than other peaks in the spectrum. This set of modulations develops a large number of peaks so to facilitate comparison and presentation we calculate a 1460 day (four year) sequence using the modulation relation outlined below and present the calculated frequency spectrum of the modulation relation in Figure 10 for comparison with the observed frequency spectrum.

The modulation relation is calculated using

$y = \sin(2\pi t/27) + \sin(4\pi t/27) + \sin(6\pi t/27) + \sin(8\pi t/27)$ as the sum of the harmonics of solar rotation and $merc = (\sin(2\pi t/(2*88)))^4$, $annual = 1 + \sin(2\pi t/365)$ and $equi = abs(\sin(2\pi t/365))$ as the amplitude modulations, respectively, at the period of Mercury, annually and equinoctially. The modulated relation is the product $z = merc*annual*equi*y$ and the frequency spectrum of z is given in Figure 10. We see in the spectrum of the modulation relation in Figure 10 the broadened 27 day and 13.5 day harmonic peaks around $f_1$ and $f_2$. Each of the harmonic peaks has sidebands at $f_n$ +/- $f_M$ due to modulation at $T_M$ = 88 days and each of these sidebands is further split by modulation at 365 days. The splitting results in about thirty peaks that go towards forming the broad band of peaks across the frequency range illustrated in Figure 10. There may be further types of splitting as each of the observed lower frequency peaks in Figure 10, those at frequencies less than 0.02 days$^{-1}$, may represent a mechanism that also modulates each of the solar harmonic variations. The overall effect of multiple modulations of the type discussed is the observed spectrum in Figure 10. However, of particular interest in this paper is the attribution of the peaks at 0.026 days$^{-1}$ and 0.048 days$^{-1}$ to modulation of the effects of 27 day solar rotation at the period of Mercury. That is, we infer that active areas on the Sun are modulated at 88 day periodicity and this leads to sidebands of the solar rotation peaks, the most prominent of which, at $f_1 - f_M$, is evident at ~ 0.026 days$^{-1}$ in Figure 10.

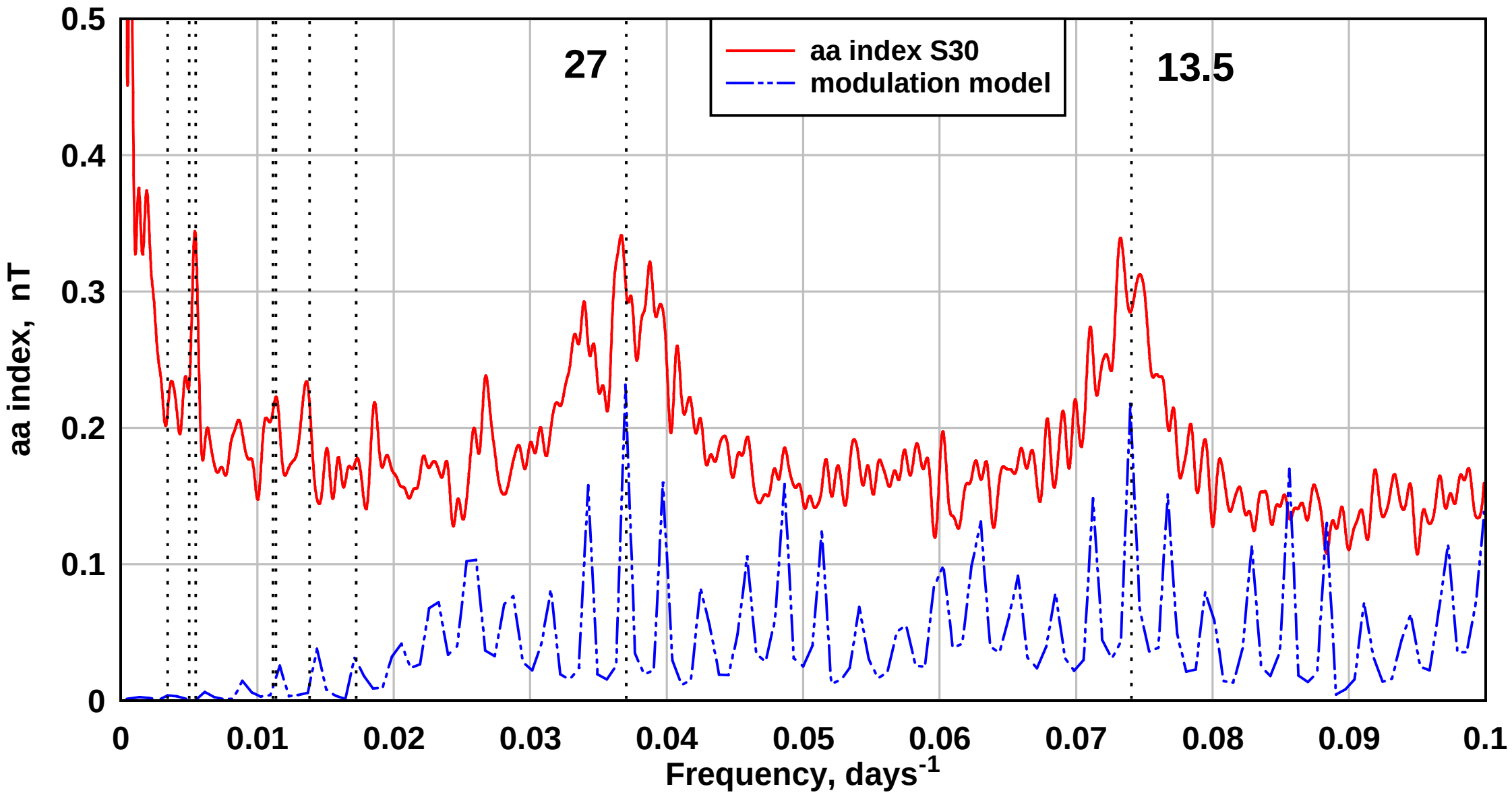

**Figure 10.** Compares the observed frequency spectrum of the aa index 1868 - 2012 and the frequency spectrum of the modulation model relation discussed in the text.

### 5.3 Prediction of long term geomagnetic activity.

Planetary orbital motion is clocklike so any effect due to planetary motion can be forecast indefinitely into the future or hind cast indefinitely into the past. The data for orbital radii on the NASA Cohoweb site can be used to derive the following expressions for the radii of Jupiter, $R_J$, and Mercury, $R_M$, as a function of time, in days, relative to January 01, 1995; $R_J = (-0.25)\cos(2\pi(t-1595)/4332.820) + 5.203$ A.U. and $R_M = (-0.07975)\cos(2\pi(t-$ 24.5)/87.96926) +0.38725 A.U. From the radii the rectified tidal effect can be found as outlined in Section 3. Figure 11 shows the predicted tidal effect for the interval between 2010 and 2057. We note that 2013 should be the first year that exhibits predominant quad-annual variation in the ~13.5 day component of aa index. There are seven intervals of quad-annual variation expected during 2010 - 2057 with the strongest quad-annual effect in 2046. We can also identify years when the semi-annual effect is predominant and symmetric, similar to the semi-annual variation in the years 1954 and 1996 selected by Cliver et al (2004). We can also identify years having a highly asymmetric semi-annual effect e.g. 2012 and 2021.

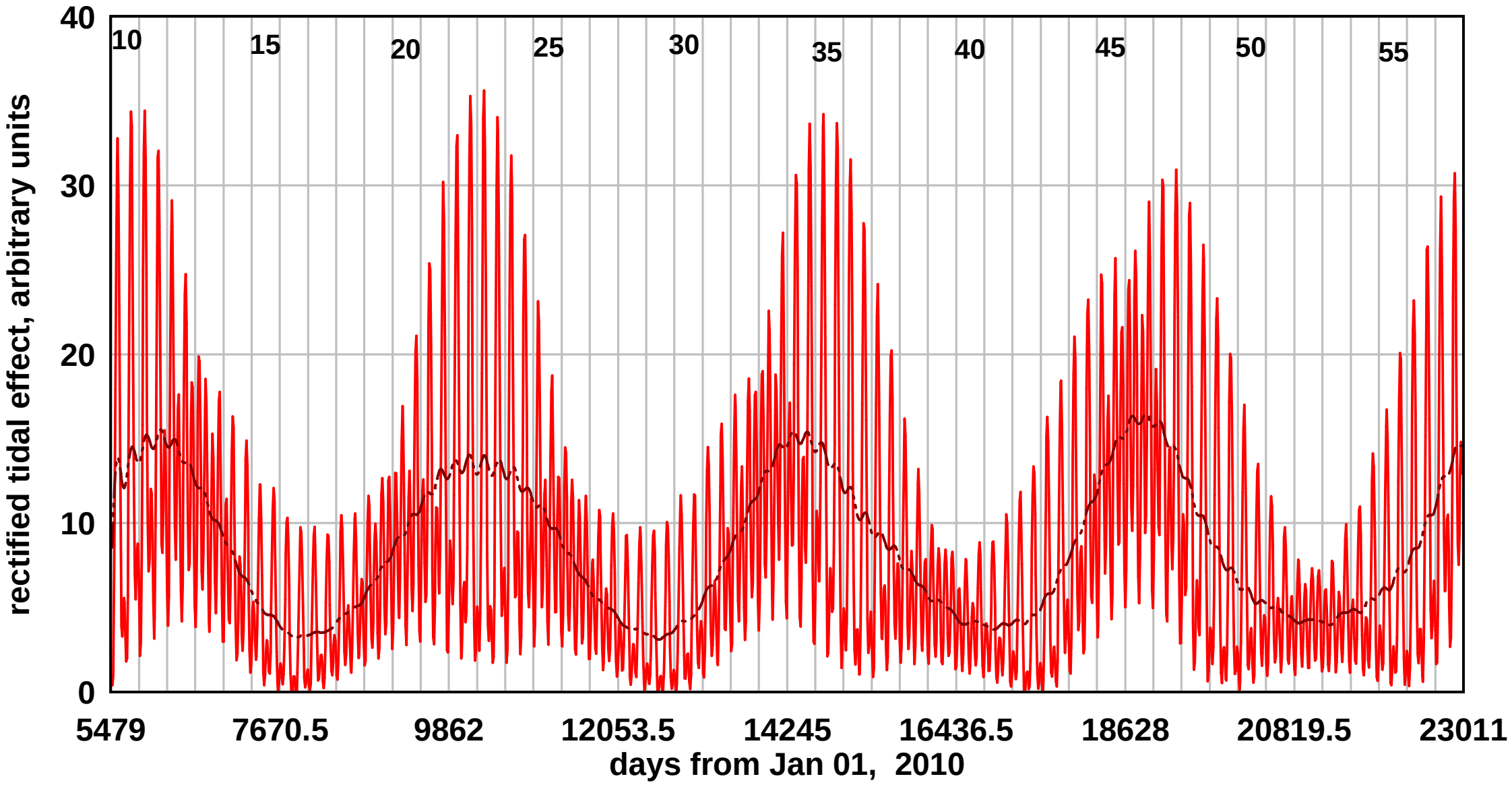

**Figure 11.** The rectified tidal effect predicted for years 2010 to 2057.

The time variation of the model over an extended interval as illustrated in Figure 6 or in Figure 11 provides an indication of the long term variation of a component of geomagnetic activity. Effectively the model predicts the variation of a component of solar output due to a tidal effect and the input of the energy from the Sun into Earth's magnetosphere. An interesting feature of the model is that the predicted maximums in geomagnetic activity, which occur at about 12 year intervals, do not always coincide with the solar cycle maxima. For example, Figure 6 shows that the model maxima in 1988 and 1999 are fairly close in time to the maxima of solar cycles 22 and 23. However, the model maximum in 1975 lies midway between the maxima of solar cycle 20 (~1968) and solar cycle 21 (~1979) and close to the solar cycle minimum (~1975). So the model predicts maximum energy associated with the tidal effect being input into the magnetosphere at 1975 and near solar cycle minimum. This at first sight seems counterintuitive however it is consistent with the observation, (Figure 12, Tsurutani et al (2006)), of the time distribution of the rate of occurrence of magnetic storms during the interval 1958 to 2004. Tsurutani et al (2006) observed that "it is found that there was far more energy injected during the 1974 declining phase of the solar cycle than during solar maxima for years 1979 and 1981". Thus the observations by Tsurutani et al (2006) provide support for the model and also support the possibility that the model may be useful in predicting future solar and geomagnetic activity.

### 5.4 Hind casting long term geomagnetic activity.

The tidal effect model can be hind cast indefinitely into the past. The aa index extends back to 1868 so there are a further 100 years of data that can be analyzed in a similar way to the analysis in Section 4. However, in the interests of a concise paper we examine only one short interval. Figure 12 shows the rectified tidal effect for years 1925 to 1964. A predominantly quad-annual effect occurs in 1953 1954 and as the two years are near solar cycle minimum and the quad-annual effect is strong the two years are favorable for analysis.

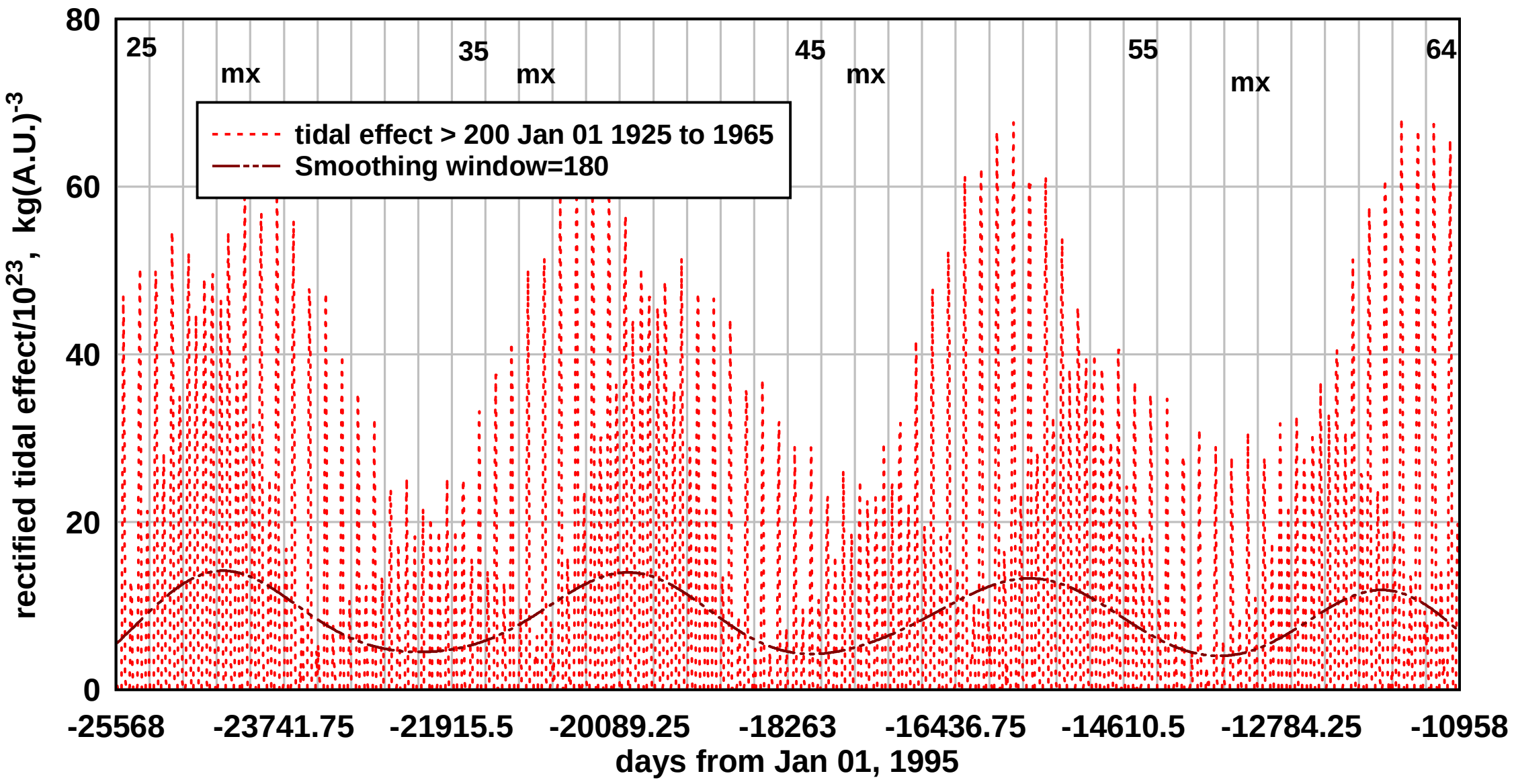

**Figure 12.** The rectified tidal effect hind cast from Jan 01 1925 to Jan 01, 1965.

Figure 13A shows the amplitude of the ~13.5 day component of the aa index for 1953 1954. A quad-annual variation is evident in both years. Figure 13B shows the frequency spectrum of the amplitude of the ~13.5 day component of aa index for the two years and the frequency spectrum of the model rectified tidal effect for the same two years, scaled to fit. The frequency resolution for spectra over a two year interval is low. However, it is clear that the dominant peak in both spectrums is near the 88 day period of Mercury and in each spectrum there are peaks near 182 days (0.0055 days$^{-1}$) corresponding to the semi-annual component and peaks near 60 days (0.0167 days$^{-1}$) corresponding to a harmonic of the semi-annual variation. If a spectrum is taken over the 15,000 day data of Figure 12 significant components of the model are found at 182.7 days (6 units), 88.0 days (14 units), 59.5 days (4 units) and 44.0 days (5 units) indicative of the longer term average ratio of quad-annual variation to semi-annual variation in the model. Clearly spectra taken over intervals of a few years are expected to vary between being dominated by the quad-annual effect with the major peak at 88 days as above or being dominated by the semi-annual effect with the major peak at 182 days. However, this is not examined further in this paper.

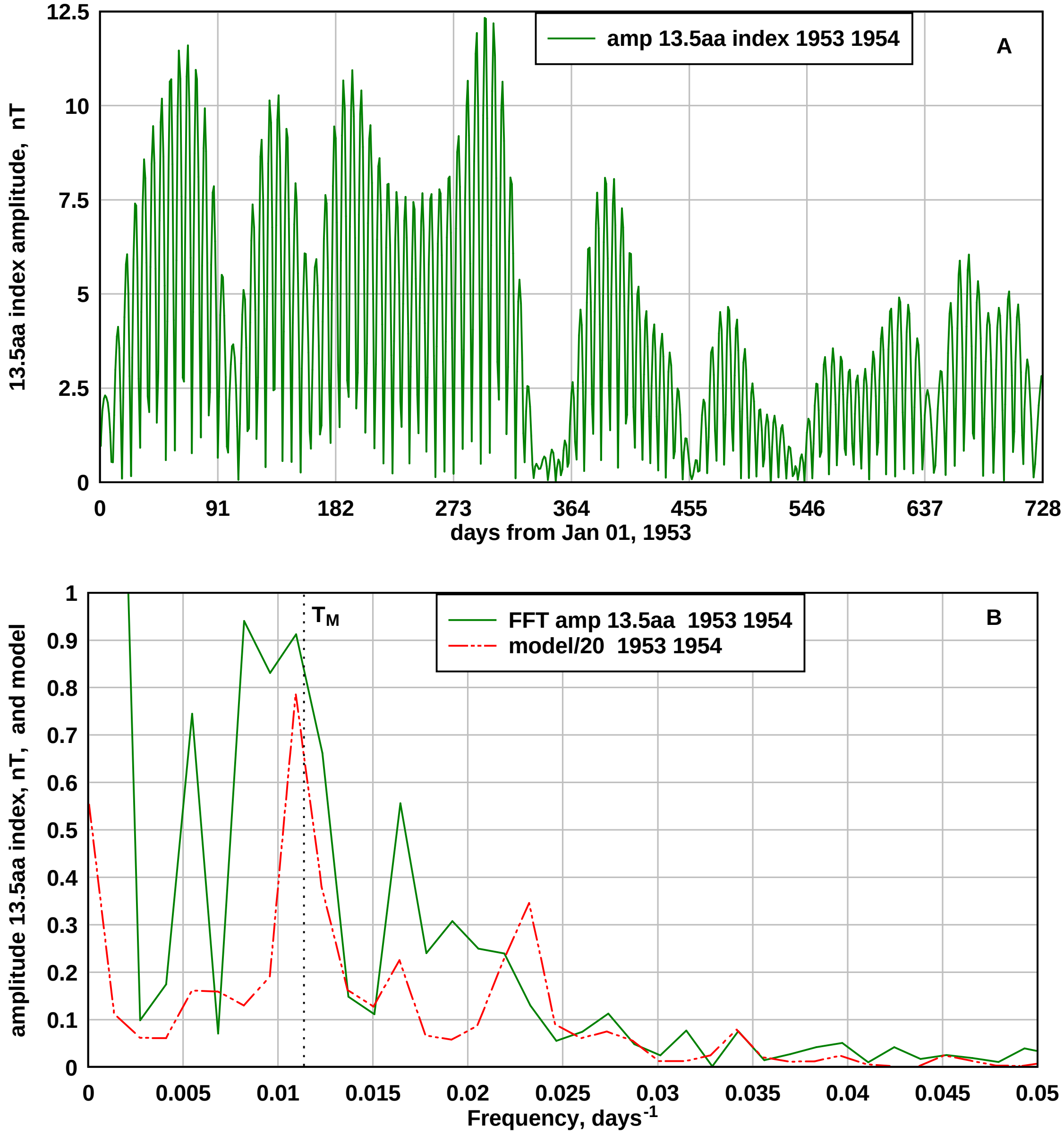

**Figure 13.** (A) The amplitude of the ~13.5 day of the daily aa index 1953 1954. (B) The spectrum of the amplitude of the 13.5 day component of aa index compared with the spectrum of the daily tidal effect for 1953 1954.

## 5.5 Tidal effect on solar cycle amplitude.

On of the intriguing aspects of this work is the observation in Section 5.3 that the broad maxima in the tidal effect that occur at intervals of about twelve years do not always coincide with solar cycle maxima and occasionally, as in 1975, Figure 6, occur at solar cycle minima. In section 5.3 it was pointed out that this was consistent with observations of Tsurutani et al (2006) that more energy was injected into the magnetosphere at the decline towards this solar cycle minimum than during the following solar cycle maximum. We infer that this may be the reason for the reduced amplitude of the prior solar cycle 20. More explicitly we suggest that when the tidal effect on solar activity is in anti-phase with the solar dynamo effect on solar activity, Charbonneau (2010), the solar activity, as measured by sunspot number at the peak of the solar cycle, is reduced. To test this idea we hind cast the tidal effect model back to the Dalton Minimum period, solar cycles 5, 6 and 7, when sunspot maximums were about one third as large as the preceding and succeeding maxima, Usoskin (2013). Figure 14 shows the tidal effect during the interval 1790 to 1830. We show only the tidal effect without modulation by the equinoctial rectification as we are here interested in solar activity on the Sun rather than geomagnetic activity at Earth. The full line in Figure 14 represents the 180 day running average of the tidal effect. It is evident that during the three solar cycles of the Dalton Minimum the three maxima of the tidal effect are in anti-phase with the three solar cycle maxima. This suggests that, similar to the case in 1975, more magnetic energy was emitted via the tidal effect around solar minima and less emitted via sunspot activity around solar maxima during this interval resulting in reduced sunspot number maximums.

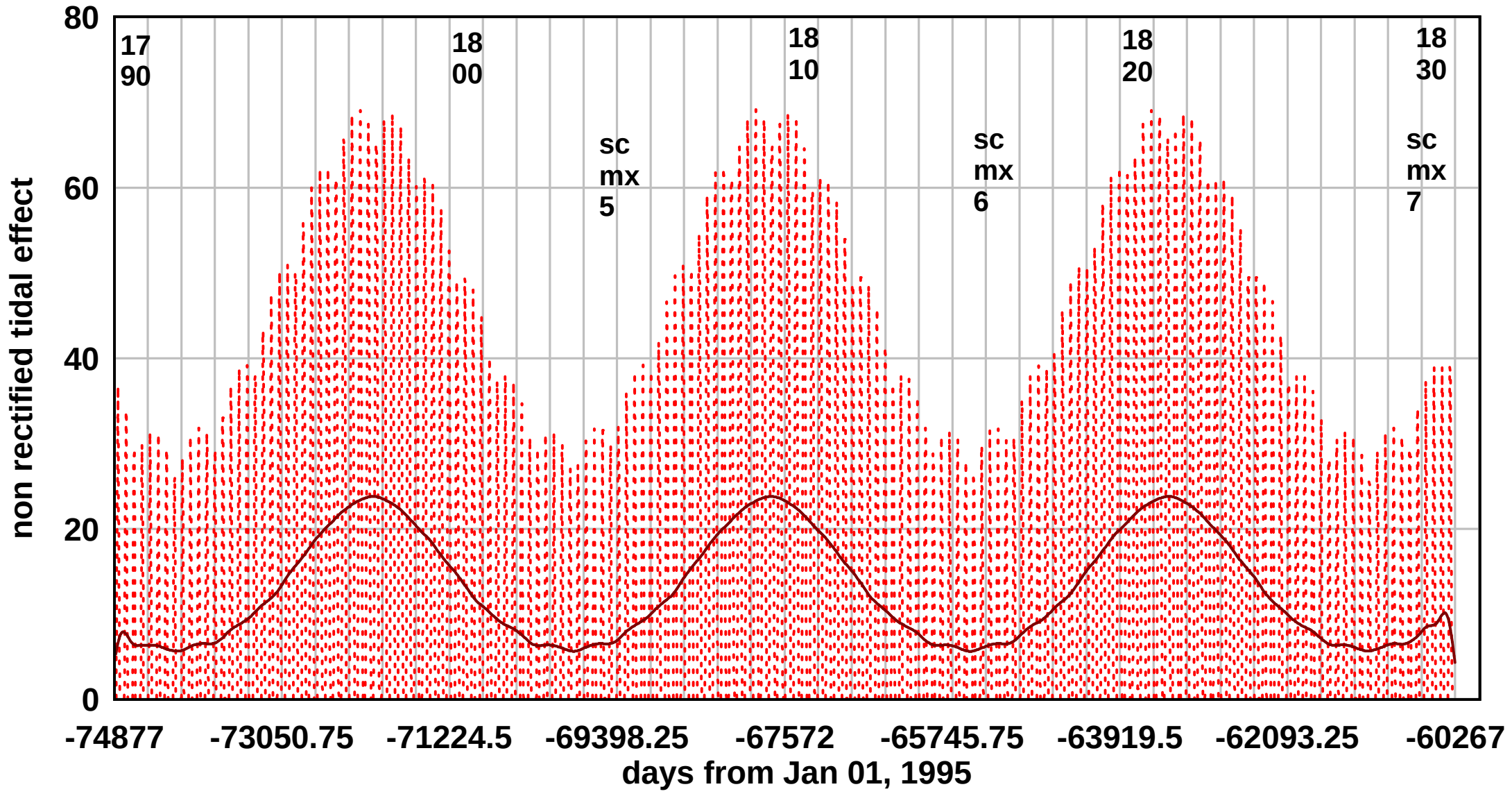

**Figure 14.** The tidal effect during the Dalton Minimum years, 1790 to1830. Also indicated are the approximate years of occurrence of the solar cycle maximums.

## 6. Conclusions

We conclude that a model based on the combined tidal effect of Mercury and Jupiter on the Sun and modified by an equinoctial effect at Earth can explain and accurately predict the intermittent occurrence of a quad-annual effect in the ~13.5 day component of the geomagnetic aa index. The model is relatively simple, based on just the variation of the orbital radii of Mercury and Jupiter via the tidal effect combination $[M_M/R_M^3 + M_J/R_J^3]$ whereas other studies of the tidal effect on the Sun have focused more on planetary conjunctions, Hung (2002), Scafetta (2012). Support for the model comes from the fact that the model accurately predicts the years when the quad-annual effect in the ~13.5 day component becomes predominant over the more frequent semi-annual effect in the ~13.5 day component. The model differs from the various prior mechanisms that attempt to explain the semi-annual effect in geomagnetic activity in that, in this model, both the quad-annual and semi-annual variation in the ~13.5 day component of geomagnetic activity derive, principally, from an 88 day period, planet Mercury, tidal effect on the Sun and the resultant 88 day period variation in solar activity. The observation of a significant near 88 day component in solar wind velocity during times of predominant quad-annual variation provided further support for a tidal effect due to Mercury. Further support for an 88 day modulation of geomagnetic activity was provided by identifying, in the spectrum of aa index, a significant peak at 88 days period and identifying, in the spectrum, significant sidebands associated with an 88 day modulation of the 27 day solar rotation period. In particular the significant peak at 39 day period was associated with the latter modulation. By comparing the standard deviations of the variations of the aa index and the ~13.5 day component of aa index in Figure 1 we estimate about 20% of the overall aa index variation may due to this tidal effect. Further support for the model comes from the ability of the model to explain and predict the occasional occurrence of high geomagnetic activity during solar cycle minimum. Other explanations for the aforementioned observations may be found. However, in the absence of other explanations, the possibility of a planetary tidal effect on solar activity and geomagnetic activity should be considered in further studies.

**References.**